\documentclass[letterpaper,10pt]{JHEP3}

\usepackage{amsfonts}
\usepackage{amssymb}
\usepackage{epsfig}
\usepackage{graphicx}
\usepackage{amsmath}
\usepackage{amsxtra}

\def\circa#1{\,\raise.3ex\hbox{$#1$\kern-.75em\lower1ex\hbox{$\sim$}}\,}
\makeatletter

\def\art{\@ifnextchar[{\eart}{\oart}}
\def\eart[#1]#2#3#4#5#6{{\rm #2}, {\em #3  #4} {\rm (#6) #5} ({\em #1})}
\def\hepart[#1]#2{{\rm #2, \em#1}}
\newcommand{\oart}[5]{{\rm #1}, {\em #2  #3} {\rm (#5) #4}}

\def\thealphaequation{\theequation\hbox to
0.6em{\hfil\alph{alphaequation}\hfil}}
\def\eqnsystem#1{
\def\@eqnnum{{\rm (\thealphaequation)}}
\def\@@eqncr{\let\@tempa\relax \ifcase\@eqcnt \def\@tempa{& & &} \or
  \def\@tempa{& &}\or \def\@tempa{&}\fi\@tempa
  \if@eqnsw\@eqnnum\refstepcounter{alphaequation}\fi
\global\@eqnswtrue\global\@eqcnt=0\cr}
\refstepcounter{equation} \let\@currentlabel\theequation \def\@tempb{#1}
\ifx\@tempb\empty\else\label{#1}\fi
\refstepcounter{alphaequation}
\let\@currentlabel\thealphaequation
\global\@eqnswtrue\global\@eqcnt=0 \tabskip\@centering\let\\=\@eqncr
$$\halign to \displaywidth\bgroup \@eqnsel\hskip\@centering
$\displaystyle\tabskip\z@{##}$&\global\@eqcnt\@ne
\hskip2\arraycolsep\hfil${##}$\hfil& \global\@eqcnt\tw@\hskip2\arraycolsep
$\displaystyle\tabskip\z@{##}$\hfil
\tabskip\@centering&\llap{##}\tabskip\z@\cr}
\def\endeqnsystem{\@@eqncr\egroup$$\global\@ignoretrue} \makeatother

\newcounter{alphaequation}[equation]

\def\beq{\begin{equation}}
\def\eeq{\end{equation}}
\def\baq{\begin{eqnarray}}
\def\eaq{\end{eqnarray}}

\def\k{{\bf k}}
\def\q{{\bf q}}

\def\x{{\bf x}}
\def\G{{\rm G}}
\def\d {{\rm d}}

\def\be{\begin{equation}}
\def\ee{\end{equation}}

\def\bea{\begin{eqnarray}}
\def\eea{\end{eqnarray}}

\def\fnl{f_{\rm NL}}
\def\nfnl{n_{f_{\rm NL}}}
\def\gnl{g_{\rm NL}}
\def\ngnl{n_{g_{\rm NL}}}
\def\taunl{\tau_{\rm NL}}
\def\ntaunl{n_{\tau_{\rm NL}}}

\def\bk{{\bf k}}

\def\fsig{f_{\sigma}}
\def\gsig{g_{\sigma}}

\preprint{BI-TP 2010/19 \\ HD-THEP-10-15 \\ YITP10-27}

\title{Scale-dependent non-Gaussianity probes inflationary
physics}

\author{C.~T.~Byrnes,$^{\,a}$ M.~Gerstenlauer,$^{\,b\,}$
  S.~Nurmi,$^{\,b\,}$
  G.~Tasinato,$^{\,b\,}$
 D. Wands$^{\,c\,d\,}$\\

$^a$ Fakult{\"a}t f{\"u}r Physik, Universit{\"a}t Bielefeld,
Postfach 100131, 33501 Bielefeld, Germany  \\
$^b$ Institut f\"ur Theoretische Physik, Universit\"at
Heidelberg, Philosophenweg 16 and 19, Germany\\
$^c$Institute of Cosmology $\&$ Gravitation, University of Portsmouth, Dennis Sciama Building,\\\hskip0.2cm Portsmouth, PO1 3FX, United Kingdom\\ $^d$ Yukawa Institute for Theoretical Physics, Kyoto University, Kyoto 606-8502, Japan
 \\ 
E-mail: \email{byrnes@physik.uni-bielefeld.de}, \email{m.gerstenlauer@thphys.uni-heidelberg.de}, \email{s.nurmi@thphys.uni-heidelberg.de}, \email{g.tasinato@thphys.uni-heidelberg.de}, \email{david.wands@port.ac.uk}}

\abstract {We calculate the scale dependence of the bispectrum and
trispectrum  in (quasi) local models of non-Gaussian
 primordial density perturbations, and characterize
  this scale dependence in terms of
  new observable parameters. They can help to discriminate between
models of inflation, since they are sensitive to properties of the
inflationary physics that are not probed by the standard observables.
  We find consistency relations between
these parameters in certain classes of models. We apply our results to
a scenario of
 modulated
reheating, showing that the scale dependence
of non-Gaussianity
 can be significant. We
also discuss the scale dependence of the bispectrum and trispectrum,
 in cases where
one varies the shape as well as the overall scale of the figure
under consideration.
We conclude providing a
formulation of the curvature perturbation in real space, which
generalises the standard local form by dropping the assumption that
$\fnl$ and $\gnl$ are constants. }

\begin{document}


\section{Introduction}

Inflation is the simplest framework  which explains
 the origin of the observed power spectrum
of temperature fluctuations in the cosmic microwave
background \cite{Komatsu:2010fb}. It is now widely accepted
 that
 non-Gaussianity is a powerful probe  to discriminate between the many
currently viable inflationary models \cite{Chen:2010xk,Byrnes:2010em,Komatsu:2010hc,Wands:2010af,Yadav:2010fz, Fergusson:2010dm,Desjacques:2010nn}. It
is usually parameterized in terms of
a single constant parameter, $\fnl$, corresponding
to the amplitude of the
bispectrum normalized to the square of the power spectrum
of primordial curvature fluctuations \cite{Salopek:1990jq,Komatsu:2001rj}. More recently it has become common to further characterize
local non-Gaussianity
including
the two non-linearity parameters associated
to the trispectrum, called  $\gnl$ and $\taunl$, again treating them as constants \cite{Byrnes:2006vq}.
 However, it has been recently pointed out,
   both from  theoretical \cite{Byrnes:2008zy,Kumar:2009ge,Byrnes:2009pe,Bernardeau:2010ay} and
observational viewpoints \cite{Sefusatti:2009xu}, that
 $\fnl$ is not necessarily constant. We show the same holds true for $\gnl$ and $\taunl$.
As happens with the power spectrum and the spectral index,
 they are characterized
by  a scale dependence,
that which denote respectively with $\nfnl$, $\ngnl$ and $\ntaunl$. 
For example, if $\fnl$ is large and positive on large scale structure scales
\cite{Jimenez:2009us,Holz:2010ck,Cayon:2010mq,Baldi:2010td}, but has a smaller value on the largest CMB scales then this would require that $\fnl$ is scale dependent.
Any scale dependence of the non-linearity parameters provides a new and potentially powerful
observational probe of inflationary physics. 





In this work,
we discuss a new approach to study the scale dependence of the
non-linearity parameters, based on
the  $\delta N$-formalism \cite{starob85,Sasaki:1995aw,Lyth:2005fi}. This
allows us to obtain an expression for the curvature perturbation
 that generalizes the local Ansatz introduced in \cite{Salopek:1990jq,Komatsu:2001rj}, and
that contains the aforementioned scale-dependent parameters.
For the single field case, the curvature
perturbation can be schematically written as
 \be
 \nonumber
  \zeta_{\k}\,=\, \zeta_{\k}^{\G}+\frac35 \fnl^{\rm p}
\left(1 +\nfnl \,{\rm ln}\,k\right)
 (\zeta^{\G}\star\zeta^{\G})_{\k}
+  \frac{9}{25} \gnl^{\rm p} \left(1 + \ngnl \,{\rm
  ln}\,k\right)(\zeta^{\G}\star\zeta^{\G}\star\zeta^{\G})_{\k}\ ,
  \ee
 where $\zeta^{\G}_{\k}$ is a Gaussian variable, and $\fnl^{\rm p}$ and $\gnl^{\rm p}$
are constants. Our approach allows us to directly calculate
  $\nfnl$, $\ngnl$ and $\ntaunl$
in models with an arbitrary
inflationary potential and an arbitrary number of fields,  assuming  slow-roll inflation. We also assume the field
perturbations are Gaussian at Hubble exit.
 Our results depend only
on the slow-roll parameters evaluated at Hubble exit, and on the
derivatives of $N$, the e-folding number. In particular, we
find that $\nfnl$ and $\ngnl$ are sensitive to third and fourth
derivatives of the potential along the directions in field space
that are responsible for generating non-Gaussianities. These do not
in general coincide with the adiabatic direction (during inflation) and such features
cannot therefore be probed by only studying the spectral index and
its running \cite{Wands:2003pw}. In the case that a single field
generates the curvature perturbation there is a consistency relation
between $\nfnl$ and $\ntaunl$ which is the derivative of the
consistency relation between $\fnl$ and $\taunl$. We explicitly show
how this consistency relation is violated in multiple field models.

In the framework of slow-roll inflation, there are
various ways to generate large non-Gaussianity, in models
in which more than one field play a role during the
inflationary process. This is the case of multiple field
inflation \cite{Rigopoulos:2005us,Vernizzi:2006ve,Battefeld:2006sz,Seery:2006js,Barnaby:2006km,Choi:2007su,Yokoyama:2007uu,Byrnes:2008wi,Byrnes:2009qy,Battefeld:2009ym}, in which two or more fields
contribute to the curvature perturbations. But there
are also approaches
in which, although more than one field is
light during inflation, only one of them
contributes  significantly  to the curvature perturbations
 (the most studied examples are the curvaton \cite{curvaton}
and modulated reheating \cite{modulatedreheating,Suyama:2007bg} scenarios).  In this work,
  we apply our general  findings  to this last class of models.
  We consider
set-ups in which  an isocurvature field remains subdominant during
inflation (as required in order to have an observable level of
non-Gaussianity \cite{Langlois:2008vk}), but represents the main
source of curvature fluctuations after inflation ends. In this case,
neither the spectral index of the power spectrum of curvature
perturbations, nor its running are sensitive to the third and fourth
derivative of the potential of the subdominant field. Hence the scale dependence of non-Gaussian
 parameters
 provide a unique opportunity to probe
self interactions in these scenarios.
 As an example, in the
 modulated reheating scenario, it is possible
   for any of the non-Gaussian  parameters to
be large. We show that if the modulaton field has self interactions,
for example a quartic potential, then all of $\fnl$, $\nfnl$, $\gnl$ and
$\ngnl$ can be large and
 provide novel information about the mechanism which generates
curvature perturbations. We will also consider mixed scenarios in which the inflaton perturbations are not neglected \cite{Langlois:2004nn,Lazarides:2004we,Ichikawa:2008iq,Ichikawa:2008ne}. We note that the scale dependence of equilateral type non-Gaussianity is also of theoretical and observational interest \cite{Chen:2005fe,LoVerde:2007ri,Khoury:2008wj,RenauxPetel:2009sj}.

We have previously shown \cite{Byrnes:2009pe} that provided one
scales all three sides of the bispectrum at the same rate then
$\nfnl$ is a constant (and hence it is simplest to focus on an
equilateral configuration). We show a similar result for the
trispectrum parameters. Since it may be of interest to consider
more general variations in which one changes the shape  of the
figure under consideration, we also consider this case.
 We
find the combination of shape and scale dependence which maximizes
$\nfnl$ and show that it is never significantly larger than the
standard result, in which one keeps the shape fixed. However we single
out
interesting limits in which there is no scale dependence, corresponding
to squeezed figures.

While in most of the paper we work in momentum space, in the last
part we also discuss how to
 describe our results in coordinate space. We provide an expression for
the curvature perturbations in
real space, that generalizes
the simplest local Ansatz \cite{Komatsu:2001rj}, and that exhibits
directly in coordinate space the effect
of  scale dependence of non-Gaussian parameters.

The plan of our paper is as follows: In Sec.~\ref{sec:general} we
extend and simplify the results from our previous paper to give
general results for the non-linearity parameters, including those
which measure the trispectrum. In Sec.~\ref{sec:onefield} we reduce
the results to general single field models and derive a consistency
relation. In Sec.~\ref{sec:twofield} we consider simple one or
two-field models in which one field, e.g.~ the curvaton, generates
non-Gaussianity but we do not exclude the Gaussian perturbations
from the inflaton. Many popular models in the literature fall into
this class and the reader may choose to skip straight to this
section where the results and notation are significantly simpler. As
an explicit example we study modulated reheating. In
Sec.~\ref{sec:shapedependence} we consider in detail the scale
dependence
 of the bispectrum and the trispectrum, and how this can be affected by the
shape of the triangle or quadrilaterum. In Sec.~\ref{sec:coordspace} we consider how to
generalize the coordinate space expression of the curvature
perturbation to include scale dependence. Finally we conclude in
Sec.~\ref{sec:conclusions}.

\section{General results}\label{sec:general}

In this section, we discuss a new approach
 to analyze the scale dependence of quasi-local non-Gaussianity,
by means of a suitable implementation of the $\delta N$ formalism.
Using the $\delta N$ formalism \cite{starob85,Sasaki:1995aw,Lyth:2005fi}, the curvature perturbation for a
system of $n$ scalar fields $\phi_{a}$ is given by the expression
  \beq
  \label{zeta_ti}
  \zeta(t_{f},\x)=\sum_{a}N_{a}(t_{f},t_{i})\delta\phi^{a}(t_{i},\x)
  +\frac{1}{2}\sum_{ab}N_{ab}(t_f,t_i)
  \delta\phi^{a}(t_i,\x) \delta\phi^{b}(t_i,\x)+\cdots\
  ,
  \eeq
where $t_f$ labels a uniform energy density hypersurface and $t_i$
denotes a spatially flat hypersurface. The result is valid on
super-horizon scales where spatial gradients can be neglected.
  In this work, we do not consider secondary effects
 on curvature perturbations arising
 from late-time physics (see \cite{Bartolo:2010qu} for a review),
nor the effects of possible isocurvature modes during the late universe.
 The
quantities $N_a$ and $N_{ab}$ denote derivatives of the number of
e-foldings along the scalar fields. We choose $t_i$ as a time soon
after the horizon crossing of all the modes of interest. Written in
momentum space, Eq.~(\ref{zeta_ti}) reads
  \beq
  \label{zeta_ti_fourier}
  \zeta_{\k} (t_f)=\sum_{a}N_{a}(t_f,t_i)\delta\phi^{a}_{\k} (t_i)
  +\frac{1}{2}\sum_{ab}N_{ab}(t_f,t_i)
  \left(\delta\phi^{a}(t_i)\star \delta\phi^{b}(t_i)\right)_{\k}
  +\cdots\ .
  \eeq
Here $k<a(t_i)H(t_i)$, since we focus on super-horizon scales, and
$\star$ denotes a convolution:
  \beq
  \left(\delta\phi^a(t_i)\star\delta\phi^b(t_i)\right)_{\k}\equiv\int
  \frac{{\rm d}{\bf q}}{(2\pi)^3}\,\delta\phi^a_{{\bf q}}(t_i)\delta\phi^b_{\k-{\bf q}}(t_i)\
  .
  \eeq

To analyze the statistical properties of the curvature perturbation
it is useful to express the results in terms of scalar perturbations
evaluated at horizon crossing $\delta\phi^{a}_{\k} (t_{k})$.
Assuming the fields $\phi_{a}$ obey slow roll dynamics during
inflation and have canonical kinetic terms, $\delta\phi^{a}_{\k}
(t_{k})$ are Gaussian up to slow roll corrections
\cite{Maldacena:2002vr,Seery:2005gb}. These corrections are
irrelevant in cases where the non-Gaussianities are large,
$|\fnl|\gg 1$ or $|\gnl|\gg 1$. Therefore, in our analysis we take
the fields $\delta\phi^{a}_{\k} (t_{k})$ to be Gaussian at horizon
crossing, $k=a(t_k)H(t_k)$. In \cite{Byrnes:2009pe} the result
(\ref{zeta_ti_fourier}) was expressed in terms of
$\delta\phi^{a}_{\k} (t_{k})$ by setting $t_i \rightarrow t_{k}(k)$.
This makes the coefficients $N_{ab...}$ implicitly dependent on $k$.
In this work, we follow a different approach, choosing a fixed $t_i$
for all observable $k$ modes, and explicitly solving for the
perturbations at $t_i$ as a function of $\delta\phi^{a}_{\k}
(t_{k})$. Besides being more transparent, this method allows us to
easily write down explicit results for the scale-dependence of
non-linearity parameters. The two approaches are compared in detail
in Appendix \ref{App:B}.

We first note that, assuming slow roll, the evolution of
super-horizon scale fluctuations $\delta\phi^a(t,\x)$ from some
initial spatially flat hypersurface at $t_{0}<t_{i}$ to the
spatially flat hypersurface at $t_{i}$ can be expressed in terms of
the Taylor expansion
  \baq
  \label{phi_ti}
  \delta\phi^{a}(t_i, \x)&=&
  \sum_{b}\frac{\partial\phi^{a}(t_i)}{\partial\phi^b(t_0)}
  \delta\phi^b (t_0, \x)+
  \frac{1}{2}\sum_{bc}\frac{\partial^2\phi^{a}(t_i)}{
  \partial\phi^b(t_0)\partial\phi^c(t_0)}\delta\phi^b(t_0, \x)
  \delta\phi^c (t_0, \x)+\cdots\ .
  \eaq
Here we have also assumed the fields have canonical kinetic terms,
i.e. the metric in field space is flat. The result (\ref{phi_ti})
follows directly from the application of the $\delta N$ formalism
where any super-horizon region, labeled by $\x$, evolves as a
separate FRW universe with its own initial conditions. Since we
assume that slow roll conditions are satisfied, the initial
conditions are set by the field values $\{ \phi^a(t_0) \}$ alone,
i.e. any dependence on the field time derivatives  can be neglected.
Therefore,
  \beq
  \delta\phi^{a}(t_i,\x)=\phi^{a}(t_i)(\{\phi^b(t_{0})+\delta\phi^b(t_{0},\x)\})
  -\phi^{a}(t_i)(\{\phi^b(t_{0})\})\ ,
  \eeq
where $\phi^a(t_i)(...)$ denote FRW solutions with the initial
conditions set at $t_{0}$. Eq.~(\ref{phi_ti}) is obtained by
expanding this with respect to $\{\delta\phi^b(t_{0},\x)\}$ while
keeping fixed the number of e-foldings between $t_0$ and $t_i$. This
corresponds to choosing $t_0$ and $t_i$ as spatially flat
hyper-surfaces, since it amounts to comparing different realizations
of FRW universes that all undergo the same number of e-foldings
between $t_0$ and $t_i$.

The coefficients in Eq.~(\ref{phi_ti}) can easily be computed by
solving the slow roll equations of motion, $3 H \dot{\phi}_a = -V_a
$. We find
  \beq
  \label{phi_sr}
  \phi^a(t_i) = \phi^a(t_0)-\sqrt{2\epsilon_a}\,{\rm ln}(a_i/a_0)+{\cal O}\left(\epsilon^{3/2}{\rm ln}^2(a_i/a_0)\right)\ ,
  \eeq
where the slow roll parameters are evaluated at $t_{0}$ and defined
as usual: $\epsilon_a=(V_a/(3H^2))^2/2$ and
$\eta_{ab}=V_{ab}/(3H^2)$ (with $M_{{\rm P}}\equiv 1$). In the
following we neglect the slow-roll suppressed corrections ${\cal
O}(\epsilon^{3/2}{\rm ln}^2(a_i/a_0))$, where ${\cal
O}(\epsilon^{3/2})$ denotes terms involving powers of $\epsilon_{a}$
and $\eta_{ab}$ up to $3/2$. The validity of this approximation is
discussed in more detail below and also in Appendix \ref{App:A}.
Differentiating Eq.~(\ref{phi_sr}) once with respect to the initial
field values, and keeping ${\rm ln}(a_i/a_0)$ fixed, we find
  \beq
  \label{phi_first}
  \frac{\partial\phi^{a}(t_i)}{\partial\phi^b(t_0)}=\delta_{ab}+
  \epsilon_{ab}\,{\rm ln}(a_i/a_0)\ ,
  \eeq
where we have defined
  \beq
  \epsilon_{ab}\equiv2\sqrt{\epsilon_a\epsilon_b}-\eta_{ab}\ .
  \eeq
The higher order derivatives in Eq.~(\ref{phi_sr}) can be computed
in a similar way and the results are given in Appendix \ref{App:A}.

By substituting Eq.~(\ref{phi_ti}) into the coordinate space
expression for the curvature perturbation (\ref{zeta_ti}), taking
the Fourier transform and thereafter setting $t_0=t_{k}$, we arrive
at the result
  \beq
  \label{zeta_fg_1}
  \zeta_{\k}=\sum_{a}\zeta_{\k}^{\G,a}+\sum_{ab}f_{ab}(k)(\zeta^{\G,a}\star\zeta^{\G,b})_{\k}+
  \sum_{abc}g_{abc}(k)(\zeta^{\G,a}\star\zeta^{\G,b}\star\zeta^{\G,c})_{\k}+\cdots\
  .
  \eeq
Here $\zeta_{\k}^{\G,a}$ are Gaussian fields defined as
  \beq
  \label{zeta_G}
  \zeta^{{\rm G},a}_{\k}(t_i, t_{k})
  =\sum_{b}N_a\,\delta \phi_{\k}^{b}(t_{k})\,
  \Big[\delta_{ab}+\epsilon_{ab}\,{\rm ln}\,\frac{a_iH_i}{k}\Big]\theta(a_iH_i-k)\
.
  \eeq
The Gaussianity of this quantity follows from our assumption of the
perturbations $\delta \phi_{\k}^{a}(t_{k})$ being Gaussian at the
horizon crossing.  For brevity, from now on we
 suppress the time arguments of the
derivatives of  $N$, denoting $N_{ab...}\equiv N_{ab...}(t_f,t_i)$.
The theta function in Eq.~(\ref{zeta_G}) is included to constrain
the convolutions to only include super-horizon scales, $k < a_iH_i$. The
matrices $f_{ab}(k)$ and $g_{abc}(k)$ are given by
  \baq
  \label{f}
  f_{ab}(k)&=&\frac{1}{2}\frac{N_{ab}}{N_a N_b} +\frac12\sum_{c}
 \frac{N_c F^{(2)}_{cab}}{N_{a} N_b}\,{\ln}
  \frac{a_iH_i}{k} ,
  \\
  \label{g}
  g_{abc}(k)&=&\frac{1}{6}\frac{N_{abc}}{N_a N_b N_c}
  +\frac{1}{6N_{a} N_b N_c}\sum_{d}
  \left(3 N_{da}F^{(2)}_{dbc}+N_{d}F^{(3)}_{dabc}\right){\rm
  ln}\frac{a_iH_i}{k}\ ,
  \eaq
where $k<a_iH_i$ and $F^{(m)}_{ab_1...b_m}$ denotes the
$k$-independent part of the $m$:th order coefficient in
Eq.~(\ref{phi_ti}). They are proportional to combinations of slow
roll parameters and their explicit expressions are given in Appendix
\ref{App:A}.

Our  results  are derived to first order in ${\rm
ln}\,{a_iH_i}/{k}$. In Eqs.~(\ref{f}) and (\ref{g}) the terms
proportional to ${\rm ln}\,{a_iH_i}/{k}$ represent small corrections
to the $k$-independent  parts except in the cases where
$f_{ab}(k_i)$ and $g_{abc}(k_i)$ are comparable to slow roll
parameters or even smaller. Such components, however, do not
generate observable non-Gaussianity and therefore do not play an
important role in our discussion. For this reason, we can safely
perform the expansion in ${\rm ln}\,{a_iH_i}/{k}$. Since the higher
order terms arising in this expansion are further slow roll
suppressed, and since we can choose $t_i$  such that the logarithms
never get larger than ${\cal O}(10)$ for the super-horizon modes in
our observable universe, we can truncate the expansion at first
order.

Instead of expanding in ${\rm ln}\,{a_iH_i}/{k}$, we can also choose
one of our observable super-horizon modes as a pivot-scale, $k_{\rm
p}<a_iH_i$, and expand Eqs.~(\ref{f}) and (\ref{g}) around this
point. To first order in ${\rm ln} ({k}/k_{\rm p})$ the results are
given by
  \baq
  \label{fexp}
  f_{ab}(k)&=&f_{ab}(k_{\rm p})\left(1 +n_{f,ab}\,{\rm ln}
  \frac{k}{k_{\rm p}}\right)\ ,\\
  \label{gexp}
  g_{abc}(k)&=&g_{abc}(k_{\rm p})\left(1 +n_{g,abc}\,{\rm ln}
  \frac{k}{k_{\rm p}}\right)\ ,
  \eaq
where we have defined\footnote{ It is important to realise that
$n_{f,ab}$ $(n_{g,abc})$ is only defined in the case where
$f_{ab}\neq 0$ $(g_{abc}\neq 0)$ in the limit $k \rightarrow
a_iH_i$. If $f_{a b}$ $(g_{a b})$ vanishes, it is convenient to
define the derivative in Eq.~(\ref{dnfab}) to be identically zero.}
  \baq
  \label{dnfab}
  n_{f,ab}
  \equiv \frac{d\ln |f_{ab}|}{d\ln k} = -\sum_{c}\frac{N_c
  F^{(2)}_{cab}}{N_{ab}}\ ,
  \;\;\;\;n_{g,abc} \equiv \frac{d\ln|g_{abc}|}{d\ln k}
  = -\sum_{d}
   \left(3
   \frac{N_{da}}{N_{abc}}F^{(2)}_{dbc}+\frac{N_{d}}{N_{abc}}F^{(3)}_{dabc}\right)\
   .\;\;\;\;
  \eaq
Provided that $n_{f,ab}$ and $n_{g,abc}$ are not much larger than
${\cal O}(0.01)$, truncating the above series at first order leads
to an error of a few per cents at most.  Neglecting slow roll
corrections, we can write $f_{ab}(k_{\rm p})={N_{ab}}/(2N_a N_b)$
and $g_{abc}(k_{\rm p})=N_{abc}/(6N_a N_b N_c)$. This precision
suffices when treating the $k$-independent terms in our expressions,
since we are only interested in computing scale-dependencies to
leading order in slow roll. In what follows, we will therefore
always write the constant terms to leading order precision in slow
roll.

Finally, using Eqs.~(\ref{fexp}) and (\ref{gexp}), we can express
Eq.~(\ref{zeta_fg_1}) as
  \baq
  \label{zeta_k_3rd}
  \zeta_{\k}&=&\sum_{a}\zeta_{\k}^{\G,a}+\sum_{ab}{f}_{ab}(k_{\rm p})\left(1 +n_{f,ab}\,{\rm ln}
  \frac{k}{k_{\rm p}}\right)
  (\zeta^{\G,a}\star\zeta^{\G,b})_{\k}\\\nonumber&&
 + \sum_{abc}{g}_{abc}(k_{\rm p})\left(1 +n_{g,abc}\,{\rm ln}
  \frac{k}{k_{\rm p}}\right)(\zeta^{\G,a}\star\zeta^{\G,b}\star\zeta^{\G,c})_{\k}+\cdots\
  .
  \eaq
This result is the starting point for our analysis of the
scale-dependence of non-linearity parameters. Explicit expressions
for $n_{f,ab}$ and $n_{g,abc}$ are given in Appendix \ref{App:A} and
the scale-dependency arising from the fields $\zeta^{\G,a}_{\k}$ can
be computed using standard methods. Therefore, using
Eq.~(\ref{zeta_k_3rd}) we can explicitly compute the
scale-dependencies of $f_{\rm NL}$, $g_{\rm NL}$ and $\tau_{\rm NL}$
for any model with slow roll dynamics during inflation.

\subsection{Two point function and power spectrum}

Here we re-derive some well-known results for the scale dependence
of the spectrum of curvature perturbations; they will be useful in
what follows for analyzing the scale-dependence of bispectrum and
trispectrum. The two point function of the scalar field
perturbations $\delta\phi_{\k}^{a}(t_{k})$ at horizon crossing is
given by
  \beq
  \langle\delta\phi^a_{\k_1}(t_{k})
  \delta\phi^b_{\k_2}(t_{k})\rangle = (2\pi)^3\delta(\k_1+\k_2)\frac{2\pi^2}{k_1^3}\,
  \left(\frac{H(t_{k}(k_1))}{2\pi}\right)^{2}\left[\delta_{ab}+2c\,(1-\delta_{ab})\,\epsilon_{ab}
  \right]\ ,
  \eeq
where $c=2-\ln2-\gamma\simeq0.73$ with $\gamma$ being the
Euler-Mascheroni constant. Both the diagonal $a=b$ and off-diagonal
$a\neq b $ components are given to leading order in slow roll. Note
that although the off-diagonal components are slow roll suppressed
compared to the diagonal components, their scale dependence has no
further suppression and therefore gives a contribution comparable to
the scale-dependence of the diagonal components. Therefore, we need
to retain the off-diagonal contributions in our analysis.

Using this together with Eqs.~(\ref{zeta_G}) and (\ref{zeta_k_3rd}),
we can express the power spectrum of $\zeta$, defined by
$\langle\zeta_{\k_1}\zeta_{\k_2}\rangle \equiv
(2\pi)^3\delta(\k_1+\k_2)\,P(k_1)$, in the form
  \beq
  P(k)=\frac{2\pi^2}{k^3}
  {\cal P}(k)=
  \frac{2\pi^2}{k^3}
  \sum_{ab} {\cal P}_{ab}(k)\ .
  \eeq
Here
  \be
  \label{Pab}
  {\cal P}_{ab}(k)\equiv \left(\frac{H(t_i)}{2\pi}\right)^{2}
N_a N_b\,\left[\delta_{ab}\left(1-2 \epsilon_{\rm H}\,
  \ln{\frac{k}{k_p}} \right)
+2\epsilon_{ab}\left(\tilde{c}-\ln{\frac{k}{k_p}}\right)\right]\ ,
  \ee
and we have defined $\epsilon_{\rm H}=-\dot{H}/H^2$ and
$\tilde{c}=c+{\rm ln}(a_iH_i/k_{\rm p})$. Subleading slow roll
corrections are again neglected in the scale-independent terms.

Defining a quantity
  \beq
  \label{n_ab}
  n_{ab}-1\equiv\frac{d\ln {{\cal P}_{ab}}}{d\ln
  k}=-\delta_{ab}(2\epsilon_{\rm H}+2\epsilon_{ab})-\frac{1}{\tilde{c}}(1-\delta_{ab})\ ,
  \eeq
we can write the spectral index as
  \beq
  \label{n_zeta}
  n_{\zeta}-1\equiv\frac{d\ln {\cal P}}{d\ln k}=\sum_{ab}
\left(\frac{{\cal P}_{ab}}{\cal
  P}\right)n_{ab}-1=-2\epsilon_{\rm
  H}-2\frac{\sum_{ab}\epsilon_{ab}N_aN_b}{\sum_cN_c^2}\ .
  \eeq
This agrees with the result given in \cite{Sasaki:1995aw}.

\subsection{Three point function and $f_{\rm NL}$}\label{sec_tpf}

We now proceed to apply our approach to derive the scale dependence
of non-linearity parameters. Using previous definitions we can
write $\fnl$ in a general multiple field case as
  \bea
  f_{\rm NL}(k_1,k_2,k_3)&\equiv&
\frac{5}{6}\frac{B(k_1,k_2,k_3)}{P(k_1)P(k_2)+2\,{\rm perms}}\nonumber\\
 &=&\frac{5}{3}
  \frac{\sum_{abcd}(k_1k_2)^{-3}{\cal P}_{ac}(k_1) {\cal P}_{bd}(k_2) f_{cd}(k_3)+2\,{\rm perms}}{
  (k_1k_2)^{-3}{\cal P}(k_1){\cal P}(k_2)+2\,{\rm perms}}\label{fnlsd}\ ,
  \eea
where the bispectrum is defined by
$(2\pi)^3\delta(\k_1+\k_2+\k_3)B(k_1,k_2,k_3)=\langle\zeta_{\k_1}\zeta_{\k_2}\zeta_{\k_3}\rangle$.

In the equilateral case, $k_i=k$ for $i=1,2$ and $3$, this simplifies to
  \be
  f_{\rm NL}(k) = \frac{5}{3}\frac{\sum_{abcd}{\cal P}_{ac}(k) {\cal P}_{bd}(k)
  f_{cd}(k)}{{\cal P}(k)^2}\ ,
  \label{fnlequilateral}
  \ee
and  using Eqs.~(\ref{fexp}) and (\ref{Pab}), we find the scale
dependence of $f_{\rm NL}(k)$ is given by
  \beq
  \label{n_fnl_equil}
  \nfnl\equiv\frac{d\ln |\fnl(k)|}{d \ln k}=
  \frac{1}{\fnl(k_{\rm p})}\sum_{ab}f_{\rm NL}^{ab}\left(2n_{{\rm multi},a} +n_{f,
  ab}\right)\ .
  \eeq
Here we have defined \cite{Lyth:2005fi}
  \beq
  f_{\rm
  NL}^{ab}\equiv\frac{5}{6}\frac{N_aN_bN_{ab}}{
  (\sum_{c}N_c^2)^2}\
  ,\qquad \fnl(k_{\rm p})=\sum_{ab}f_{\rm
  NL}^{ab}\ ,
  \eeq
and
   \bea
   n_{{\rm multi},a}&\equiv&n_{aa}-n_{\zeta}-2\sum_{c}(1-\delta_{ac})\epsilon_{ac}\frac{N_c}{N_a}\\
   &=& 2\sum_{cd}\epsilon_{cd}\left(\frac{N_cN_d}{\sum_{b}N_b^2}-
   \delta_{ad}\frac{N_c}{N_a}\right)\ .
   \eea

All the quantities in Eq.~(\ref{n_fnl_equil}) depend on combinations
of slow roll parameters and on the constant coefficients $N_{a},
N_{ab}$ in the $\delta N$ expansion (recall that the explicit
expression for $n_{f,ab}$ is given by Eq.~(\ref{app_nf})). For a
given model these can all be regarded as known quantities and the
scale-dependence of $\fnl$ can therefore be directly read off from
the above result without doing any further computations.

In Eq.~(\ref{n_fnl_equil}) we can clearly identify two sources of
scale dependence. The contribution proportional to $n_{f,ab}$
follows from the non-linear evolution of perturbations outside the
horizon \cite{Byrnes:2009pe}. The part proportional to $n_{{\rm
multi},a}$ is associated with the scale dependence of factors of the
form ${\cal P}_{ac}/{\cal P}$ in equation (\ref{fnlequilateral}). It
 is present only in the multi-field case (indeed
for a single field model this factor is equal to unity) and arises due to the
presence of multiple unrelated Gaussian fields $\zeta^{\G,a}$ in the
expansion of $\zeta$ in (\ref{zeta_k_3rd}). This generates deviations
from the local form and makes $\fnl$ scale-dependent even if the
perturbations would evolve linearly outside the horizon.
 Indeed, by setting $n_{f,ab}=0$ we recover the results of a
multi-local case analyzed separately in \cite{Byrnes:2009pe}.

As shown in \cite{Byrnes:2009pe}, the scale dependence of $\fnl$ is
given by the same result (\ref{n_fnl_equil}) for the class of
variations where the sides are scaled by the same constant factor,
$\k_i\rightarrow \alpha\k_i$. For such shape-preserving variations
where only the overall scale of the triangle is varied, the result
does not depend on the triangle shape. This holds at leading order in
slow roll. Generic variations changing both the scale and the shape
of the triangle are considered in Sec. \ref{sec:shapedependence}.

\subsection{Four point function, $g_{\rm NL}$ and $\tau_{\rm NL}$}\label{sec_qpf}

The connected part of the four point correlator of $\zeta$ can be
written in the form
  \baq
  \label{zeta_4point}
  \langle\zeta_{\k_1}\zeta_{\k_2}\zeta_{\k_3}\zeta_{\k_4}\rangle
&=&(2\pi)^3\delta(\sum_{i=1}^{4}{\k_i})
  \left[\rule{0 pt}{4 ex}\right.
  \tau_{\rm NL}(k_1,k_2,k_3,k_4,k_{13})\left(\rule{0 pt}{2 ex}\right.P(k_1)P(k_2)P(|\k_1+\k_3|)
  +11\,{\rm perm}\left)\rule{0 pt}{2 ex}\right.\qquad\\\nonumber&&
  +\frac{54}{25}g_{\rm NL}(k_1,k_2,k_3,k_4)\left(\rule{0 pt}{2 ex}\right.
  P(k_1)P(k_2)P(k_3)+3\,{\rm perm}\left)\rule{0 pt}{2 ex}\right.
  \left]\rule{0 pt}{4 ex}\right.\ ,
  \eaq
where we have defined $k_{ij}\equiv |\k_i+\k_j|$. The functions
$\tau_{\rm NL}$ and $g_{\rm NL}$ are given by
  \baq
  \label{tnlsd}
  \tau_{\rm NL}(k_1,k_2,k_3,k_4,k_{13})&=&
  4\,\frac{
  (k_1 k_2 k_{13})^{-3}
  \sum_{abcdef}{\cal P}_{ac}(k_1) {\cal P}_{be}(k_2) {\cal P}_{df}(k_{13}) f_{cd}(k_3)
  f_{ef}(k_4) + 11\,{\rm perm} }{(k_1 k_2 k_{13})^{-3}\,
  {\cal P}(k_1){\cal P}(k_2){\cal P}(k_{13}) + 11\,{\rm
  perm}}\ ,\qquad\;
  \\
  g_{\rm NL}(k_1,k_2,k_3,k_4)&=&
  \frac{25}{9}
  \frac{(k_1 k_2 k_3)^{-3}
  \sum_{abcdef}P_{ad}(k_1) P_{be}(k_2) P_{cf}(k_3) g_{def}(k_4)+3\,{\rm perms}}{
  (k_1 k_2 k_3)^{-3}\,{\cal P}(k_1){\cal P}(k_2){\cal P}(k_{3}) + 3\,{\rm perm}
  }\label{gnlsd}\ .
  \eaq

In the case of a square, $k=k_i$ (notice that $\taunl$, but not $\gnl$, is sensitive to the
angles between the vectors and different equilateral figures in
general yield different results), the above expressions reduce to
  \bea
  \label{taunl_quad}
  \taunl(k)&= &
  4\sum_{abcdef}\frac{{\cal P}_{ac}(k) {\cal P}_{be}(k) {\cal P}_{df}(\sqrt{2}k) f_{cd}(k) f_{ef}(k)}{{\cal P}(k)^3}\ , \\
  \label{gnl_quad}
  \gnl(k)&=& \frac{25}{9}
  \sum_{abcdef}\frac{{\cal P}_{ad}(k) {\cal P}_{be}(k) {\cal P}_{cf}(k) g_{def}(k)}{{\cal P}(k)^3}\ .
  \eea
The scale-dependence can be computed similarly to the analysis of
the bispectrum above. Using Eqs.~(\ref{fexp}), (\ref{gexp}) and
(\ref{Pab}), we find
  \bea
  \label{ntaunl}
  \ntaunl&\equiv&\frac{d\ln |\taunl(k)|}{d \ln k}\,=\,
\frac{1}{\taunl(k_{\rm p})}\sum_{abcd}\tau_{\rm
  NL}^{abcd}\left[(2n_{\rm multi,a}
  -(n_{\zeta}-1)-2\epsilon_{\rm H})\delta_{bc}-2\epsilon_{bc}+2n_{f,ab}\delta_{bc}\right]\,\ ,\;\;\;\;\;\;\,\\
  \label{ngnl}
  \ngnl&\equiv&\frac{d\ln |\gnl(k)|}{d \ln k}=\frac{1}{\gnl(k_{\rm p})}\sum_{abc}g_{\rm NL}^{abc}\left(3
  n_{{\rm multi},a}+n_{g,abc}\right)\ ,
  \eea
where $\epsilon_H = -\dot{H}/H^2$, and \cite{Seery:2006js,Byrnes:2006vq}
  \baq
  \tau_{\rm NL}^{abcd}
  &=&\frac{N_aN_{ab}N_{cd}N_d}{(\sum_{e}N_{e}^2)^3}\ ,
\qquad \taunl(k_{\rm p})=\sum_{abcd}\tau_{\rm
  NL}^{abcd}\delta_{bc}\ ,
  \\
  g_{\rm
  NL}^{abc}&=&\frac{25}{54}\frac{N_aN_bN_cN_{abc}}{(\sum_{d}N_{d}^2)^3}
  \ , \qquad \gnl(k_{\rm p})=\sum_{abc}g_{\rm
  NL}^{abc}\ .
  \eaq
The scale-dependencies are fully determined by the constant
coefficients $N_{a}, N_{ab}, N_{abc}$ in the $\delta N$ expression
and by combinations of slow-roll parameters, which enter the results
through Eqs.~(\ref{app_nf}) and (\ref{app_ng}). Although the
expressions appear lengthy in their general form, considerable
simplifications typically occur when considering specific models. We
will discuss examples in Sections \ref{sec:onefield} and
\ref{sec:twofield}.

Similarly to $\nfnl$, we can again distinguish two physically
different contributions in the expressions for $\ntaunl$ and
$\ngnl$. The parts proportional to $n_{f,ab}$ and $n_{g,abc}$ in
Eqs.~(\ref{ntaunl}) and (\ref{ngnl}) respectively arise from the
non-linear evolution outside the horizon. The other contributions
describe deviations from the local form due to the presence of
multiple fields, similarly to what we discussed in the previous
section.

The results (\ref{ntaunl}) and (\ref{ngnl}) hold not only for the
special case of a square, but for any variations where all the sides
are scaled by the same constant factor, $\k_i\rightarrow\alpha\k_i$.
These variations preserve the shape of the momentum space figure and
change only its overall scale. We will prove this result in Sec.
\ref{sec:shapedependence} where we also discuss generic variations
that simultaneously change both the scale and the shape.

Having presenting our formalism and the general results, we will
discuss in the next two sections applications to specific cases.

\section{General single field case}\label{sec:onefield}

We start by discussing models where the primordial curvature
perturbation effectively arises from a single scalar field, which does not need to be the inflaton and we call $\sigma$.
 In this case,
 the functions $f_{\sigma\sigma}(k)$ and
$g_{\sigma\sigma\sigma}(k)$ appearing in the expansion of $\zeta$,
Eq.~(\ref{zeta_k_3rd}), are up to numerical factors equal to $f_{\rm
NL}(k)$ and $g_{\rm NL}(k)$, evaluated for the equilateral
configurations. This can be seen directly from
Eqs.~(\ref{fnlequilateral}) and (\ref{gnl_quad}). We can therefore
rewrite Eq.~(\ref{zeta_k_3rd}) as
  \be
  \label{zeta_single}
  \zeta_{\k}\,=\,
  \zeta_{\k}^{\G}+ \frac{3}{5}\fnl(k)(\zeta^{\G}\star\zeta^{\G})_{\k}+
  \frac{9}{25}\gnl(k)
  (\zeta^{\G}\star\zeta^{\G}\star\zeta^{\G})_{\k}+\cdots\ .
  \ee
As we will discuss in Sec.~\ref{sec:twofield}, this result applies
for example to the curvaton scenario and modulated reheating in the
limit where the inflaton perturbations are negligible. We therefore
call all the models where the curvature perturbation can be
expressed in the form (\ref{zeta_single}) as general single field
models.

According to Eqs.~(\ref{fexp}) and (\ref{gexp}), the non-linearity
parameters $\fnl(k)$ and $\gnl(k)$ are now given by
  \bea
  \label{fexps}
  \fnl(k)&=& \frac{5}{6}\frac{N''}{N'^2}\,\left(1+ n_{f_{\rm NL}} \,{\rm ln}
  \frac{k}{k_{\rm p}}
  \right)\ ,
  \\
  \label{gexps}
  \gnl(k)&=&\frac{25}{54}\frac{N'''}{N'^3} \left(1+ n_{g_{\rm NL}}
  \,{\rm ln}
  \frac{k}{k_{\rm p}}
  \right)\ ,
  \eea
where the primes denote derivatives with respect to $\sigma$ and
$\nfnl=n_{f,\sigma\sigma},\ngnl=n_{g,\sigma\sigma\sigma}$. Using the
explicit expressions (\ref{app_nf}) and (\ref{app_ng}) in the
Appendix \ref{App:A}, we obtain
  \baq
  \label{nfnl_single}
  \nfnl&=&\frac{N'}{N''}
\left[\sqrt{2\epsilon_{\sigma}}(4\epsilon_{\sigma}-3\eta_{\sigma\sigma})+
  \frac{V'''}{3H^2}\right]\ ,\\
  \label{ngnl_single}
  \ngnl&=&3\frac{N''^2}{N'''N'}\,\nfnl-\frac{N'}{N'''}
  \left[24\epsilon_{\sigma}^2-24\epsilon_{\sigma}\eta_{\sigma\sigma}+3\eta_{\sigma\sigma}^2+
  \frac{4\sqrt{2\epsilon_{\sigma}}\,V'''}{3H^2}-\frac{V''''}{3H^2}\right]\ .
  \eaq
The same results can of course be directly obtained from
Eqs.~(\ref{n_fnl_equil}) and (\ref{ngnl}). If $\sigma$ is an
isocurvature field during inflation, $\epsilon_{\sigma}=0$ in the
above expressions.

For the general single field case Eq.~(\ref{taunl_quad}) further
yields
  \be\label{crtnlfnl}
  \taunl(k)=\left(\frac{6 \fnl(k)}{5}\right)^2\ ,
  \ee
up to scale-independent slow roll corrections. Therefore, the
scale-dependencies of $\taunl$ and $\fnl$ are related by
  \beq\label{crscng}
  \ntaunl=2\nfnl\ .
  \eeq
This simple consistency relation is characteristic for general
single field models. In multiple field models, the relation
(\ref{crtnlfnl}) is in general violated and consequently the result
(\ref{crscng}) is no longer true.

\section{Two field  models of inflation}\label{sec:twofield}

After considering single field models, in this section we
 discuss some scenarios in which more than one field can
 play an important role in the inflationary process.
  We focus on a class of models that contains
 the most important examples of inflationary set-ups
characterized by large non-Gaussianity.

Many
 models of inflation that generate sizeable
 non-Gaussianity are characterized by the presence of a
field $\sigma$, with significant non-Gaussian perturbations, that is
isocurvature during inflation.  The inflaton field $\phi$  also has
its own perturbations, which for convenience can be considered as
Gaussian. When the inflaton
  perturbations provide non-negligible  contributions
to the curvature fluctuation spectrum, the scenario is called a
mixed scenario
\cite{Langlois:2004nn,Lazarides:2004we,Ichikawa:2008iq,Ichikawa:2008ne}.
In order to generate  large non-Gaussianity by means of the field
 $\sigma$,
  it is required that $\dot{\sigma}\ll\dot{\phi}$, and
hence $\epsilon_{\sigma}\ll\epsilon_{\phi}$ \cite{Langlois:2008vk}. From this relation, it
follows that the trajectory in field space while observable modes
exit the horizon is nearly straight. Therefore it is a  good
approximation to treat the fields as uncorrelated \cite{Byrnes:2006fr}. We also make the
common assumption that the potential is separable,
\bea W(\sigma,\phi)=U(\phi)+V(\sigma)\ . \eea Hence, the only
potentially non-negligible slow roll parameters in such a scenario
are the following
\bea \epsilon_{\rm H}=\epsilon_\phi=-\frac{\dot{H}}{H^2}\ , \qquad
\eta_{\phi}=\frac{U''}{3H^2}\ , \qquad
\eta_{\sigma}=\frac{V''}{3H^2}\ , \qquad
\xi^2_{\phi}=\frac{U'''U'}{9H^4}\ , \qquad
\xi^2_{\sigma}=\frac{V'''U'}{9H^4}\ . \eea
In this case, the curvature perturbation reads\footnote{We have used
a simplified notation for this section compared to the rest of the
paper. Since all cross terms such as $P_{\phi\sigma}$ are negligibly
small in this scenario we use only a single index $\phi$ or $\sigma$
where appropriate, e.g.~for $\eta_{\sigma}\equiv\eta_{\sigma\sigma}$ and $g_{\sigma}\equiv g_{\sigma\sigma\sigma}$.}
\bea\label{zeta2field} \zeta(\bk)=\zeta_{\bk}^{G,\phi}+\zeta_{\bk}^{G,\sigma}+
f_{\sigma}(k)\left(\zeta^{G,\sigma}\star\zeta^{G,\sigma}\right)_{\bk}+g_{\sigma}(k)(\zeta^{G,\sigma}\star\zeta^{G,\sigma}\star\zeta^{G,\sigma})_{\bk}\
.\label{cp2fc} \eea
Although the assumed form of $\zeta$ is simplified, in practice
the vast majority of models in the literature, characterized by
large quasi-local non-Gaussianity, satisfy the above Ansatz to a good enough accuracy for observational purposes. For
this reason
 we will limit our attention to models  with curvature
perturbation satisfying Eq.~(\ref{cp2fc}) in this section.

In the limit that $f_{\sigma}$ and $g_{\sigma}$ are independent of
$k$, we recover the multivariate local model \cite{Byrnes:2009pe}.
 In the case that $\zeta^{G,\phi}=0$ we have the
general single field model we have
analyzed in section \ref{sec:onefield}, but here we assume this field was an
 isocurvature mode during horizon crossing.
 We will consider
 these two cases in more detail later in this section.

The power spectrum is given by
\bea {\cal P}_{\zeta}(k)={\cal P}_{\phi}(k)+{\cal
P}_{\sigma}(k)={\cal P}_{\phi}(k) (1-w_{\sigma}(k))^{-1}\ , \eea
where we have introduced the ratio
\bea w_{\sigma}(k)=\frac{\cal{P}_{\sigma}}{\cal{P}_{\zeta}}\ .
 \eea
Note that neglecting all the slow-roll corrections, and hence also
the scale dependence,
$w_{\sigma}=N_{\sigma}^2/(N_{\phi}^2+N_{\sigma}^2)$. To lowest
order in slow roll, the spectral index $n_{\zeta}-1$ and
tensor-to-scalar ratio $r_T$ satisfy the following relations \cite{Wands:2002bn}
\bea
n_{\zeta}-1&=&(n_{\sigma}-1)w_{\sigma}+(n_{\phi}-1)(1-w_{\sigma})
\nonumber\\
&=&-(6-4w_{\sigma})\epsilon_{\rm
H}+2(1-w_{\sigma})\eta_{\phi}+2w_{\sigma}\eta_{\sigma}\ ,
\\ r_T&\equiv& \frac{{\cal P}_T}{{\cal P}_{\zeta}}
=\frac{8}{N_{\sigma}^2+N_{\phi}^2}\nonumber\\&=&8\,N_\phi^{-2}\,(1-w_{\sigma})\
, \eea
where ${\cal P}_T=8H^2_k/(4\pi^2)$ is the power spectrum of tensor
perturbations and we have defined \be n_{\sigma}-1\,=\,\frac{d\ln
{\cal P}_{\sigma}}{d\ln k}\ ,
 \qquad n_{\phi}-1=\frac{d\ln \cal{P}_{\phi}}{d\ln k}\ ,  \qquad n_{\zeta}-1=\frac{d\ln \cal{P}_{\zeta}}{d\ln k}\ .
\ee

The non-Gaussianity parameter $\fnl$ in the equilateral limit, and
the trispectrum non-linearity parameters in the case of a square
configuration, are given by
\bea \fnl(k)&=&\frac53 w_{\sigma}^2(k)f_{\sigma}(k)\ , \\
\taunl(k)&=&4w_{\sigma}(k)^2w_{\sigma}(\sqrt{2}k)\fsig^2(k)\ , \\
\gnl(k) &=&\frac{25}{9}w_{\sigma}^3(k)\gsig(k)\ . \eea
Therefore their scale dependence is given by
  \bea
  \nfnl &\equiv&
  \frac{d\ln|\fnl|}{d\ln k} =
  2(n_{\sigma}-n_{\zeta})+\frac{d\ln|f_{\sigma}|}{d\ln k}  \label{defnfnl2fm}\\
  &=& 4(1-w_{\sigma})(2\epsilon_{\rm
  H}+\eta_{\sigma}-\eta_{\phi})+\frac{N_{\sigma}}{N_{\sigma\sigma}}\left(\frac{V'''}{3 H^2}-
  \sqrt{2\epsilon_{\rm H}}\,\eta_{\sigma}\left(\frac{1}{\omega_{\sigma}}-1\right)^{1/2}\right)\
  ,\\
  \ntaunl &=& 3(n_{\sigma}-n_{\zeta})+2\frac{d\ln|f_{\sigma}|}{d\ln k}
  \\
  &=& 6(1-w_{\sigma})(2\epsilon_{\rm H}+\eta_{\sigma}-\eta_{\phi})
  +\frac{2N_{\sigma}}{N_{\sigma\sigma}}\left(\frac{V'''}{3 H^2}-
  \sqrt{2\epsilon_{\rm H}}\,\eta_{\sigma}\left(\frac{1}{\omega_{\sigma}}-1\right)^{1/2}\right)\
  ,\\
  \ngnl&=&3(n_{\sigma}-n_{\zeta})+ \frac{d\ln|\gsig|}{d\ln k}\\
  &=&6(1-w_{\sigma})(2\epsilon_{\rm H}+\eta_{\sigma}-\eta_{\phi}) +
  \frac{3\,N_{\sigma\sigma}}{N_{\sigma\sigma\sigma}}\frac{V'''}{3H^2}\\\nonumber&&
  +\frac{N_{\sigma}}{N_{\sigma\sigma\sigma}}\left(\frac{V''''}{3H^2}-3\eta_{\sigma}^2
  +\sqrt{2\epsilon_{\rm H}}\,\frac{V'''}{3H^2}\left(\frac{1}{\omega_{\sigma}}-1\right)^{1/2}\right)\ ,
  \eea
where we have used the results derived in Sec.~\ref{sec:general} and
the fact that  $N_{\phi\phi}$, $N_{\phi \sigma}$ and their
derivatives are negligible in the class of models we are
considering, see Eq.~(\ref{cp2fc}). The quantities on the right hand
side of each equation should be evaluated at an initial time $t_i$
shortly after the horizon crossing time of all the modes of
observational interest. Observe that our results for $\nfnl$ and
$\ngnl$ depend on the derivatives of the potential, in combinations
that do not correspond to traditional
 slow-roll parameters. This turns out  be useful to probe
these quantities, that cannot be tested by the power spectrum and its derivatives. We are
going to discuss  this in detail in what follows.

Observational constraints on the bispectrum are given in \cite{Komatsu:2010fb} while constraints on $\gnl$ are given in \cite{Desjacques:2009jb,Vielva:2009jz} (see also \cite{Chongchitnan:2010xz}) and for both $\gnl$ and $\taunl$ in \cite{Smidt:2010sv}. Forecasts for future constraints on all three parameters are given in \cite{Kogo:2006kh,Smidt:2010ra} while forecast constraints on $\nfnl$ are given in \cite{Sefusatti:2009xu}. There are currently no forecasts for how well the scale dependence of the trispectrum parameters could be constrained or measured. Observational constraints on a model with the form (\ref{zeta2field}), without considering the scale-dependence of $\fnl$ or $w_{\sigma}$, are given in \cite{Tseliakhovich:2010kf}.

\subsection{Limiting cases}
After presenting the general formulae for the two-field case, we
discuss
 important examples of general single
field inflation,  that arise as limiting cases of the previous
discussion of two-field inflation.

\subsubsection{Isocurvature single field}

In the case that a single field $\sigma$, which is subdominant to
the inflaton during inflation, generates the primordial curvature
perturbation, one has $w_{\sigma}=1$ which implies $n_{\sigma}=n$,
$r_T\simeq0$ and $N_{\sigma}\gg1$.

In this scenario, it is useful to express
 the spectral index and its running  up to second order,
to understand which parameters are currently constrained by
observations. From \cite{Byrnes:2006fr}, we have
\bea n_{\zeta}-1&=&-2\epsilon_{\rm
H}+2\eta_{\sigma}+\left(-\frac{22}{3}+8c\right)\epsilon_{\rm H}^2
+\frac23\eta_{\sigma}^2+\left(\frac83-4c\right)\epsilon_{\rm
H}\eta_{\phi} +
\left(\frac23-4c\right)\epsilon_{\rm H}\eta_{\sigma}\ , \\
\alpha_{\zeta}&=&-8\epsilon_{\rm H}^2+4\epsilon_{\rm
H}\eta_{\phi}+4\epsilon_{\rm H}\eta_{\sigma}\ , \eea where
$c=2-\ln2-\gamma\simeq0.73$.
 Notice that, in the previous formulae, the slow-roll parameters $\xi_\sigma$
and $\xi_\phi$ do not appear in the running of the spectral index,
because they are weighted
by negligible quantities. This implies that third and higher
derivatives of the potential do not enter in the previous quantities.

The non-Gaussianity observables (which follow as special cases of
the formulae discussed in the first part of this section,
when taking the limit $\omega_\sigma \to 1$) are
  \bea
\label{fnlgnl_pivot}
\fnl&=&\frac53f_{\sigma}=\frac56\frac{N_{\sigma\sigma}}{N_{\sigma}^2}\
, \qquad \gnl=\frac{25}{9}g_{\sigma}=
\frac{25}{54}\frac{N_{\sigma\sigma\sigma}}{N_{\sigma}^3}\ ,
\\\label{ntnlcu} \nfnl&=&\frac{\ntaunl}{2}\simeq
\frac{N_{\sigma}}{N_{\sigma\sigma}}\frac{V'''}{3H^2}\simeq
\frac56 \frac{{\rm sgn}(N_{\sigma})}{\fnl} \sqrt{\frac{r_{\rm T}}{8}}\frac{V'''}{3H^2}\ , \\
\ngnl&\simeq&
3\,\frac{N_{\sigma\sigma}}{N_{\sigma\sigma\sigma}}\frac{V'''}{3H^2}
+\frac{N_{\sigma}}{N_{\sigma\sigma\sigma}}\left(\frac{V''''}{3H^2}-3\eta_{\sigma}^2\right)\label{ngnlcu}
\\ &\simeq&
 \frac{5}{3} \frac{{\rm sgn}(N_{\sigma})\,\fnl}{g_{\rm NL}} \sqrt{\frac{r_{\rm T}}{8}}
\frac{V'''}{3H^2}+\frac{25}{54}\frac{1}{g_{\rm NL}}\frac{r_{\rm
T}}{8}\frac{V''''}{3H^2} \simeq 2\frac{\fnl^2}{g_{\rm NL}}\nfnl
+\frac{25}{54}\frac{1}{g_{\rm NL}}\frac{{\cal
P}_{\zeta}^{-1}}{6\pi^2}V''''\ \label{ngnlfinal} .
  \eea
Here $\fnl$ and $\gnl$ denote the non-linearity parameters evaluated
for equilateral configurations at some pivot scale, $k=k_{\rm p}$.
As discussed in Sec.~\ref{sec:general}, $k_{\rm p}$ can be chosen as
any of the super-horizon modes in our observable universe and the
results are independent on this choice, up to subleading slow-roll
corrections.

In the previous formulae, we have presented several different
 ways of
expressing $\nfnl$ and $\ngnl$ (in Eq.~(\ref{ngnlfinal}) we have
dropped the negligible contribution $\eta_{\sigma}^2r_{\rm
T}\lesssim 10^{-6}$). This is in order to make it easier to estimate
their magnitude in different ways, depending on the available
quantities. We also note that in some cases the previous formulae might include terms at different orders in slow roll, in which case one should neglect the subleading terms (since additional terms at the same order might also have been neglected).
 In general, they are suppressed by some combination of the
tensor-to-scalar ratio, divided by non-linearity parameters. But their
size could be significant, if $\sigma$ has either a large cubic or
quartic self interaction. As we mentioned
earlier, the power spectrum does not contain
information on these parameters, even if the running of the
spectral index can be measured. Hence $\nfnl$ appears to be the best
way of probing the cubic self interaction, while in principle
$\ngnl$ could probe the quartic derivative of the isocurvaton field.

Although we have written $\nfnl\sim1/\fnl$, the prefactor to $1/\fnl$ will in general depend on some of the same model parameters as $\fnl$ so one should not view the two parameters as being inversely proportional (an explicit example is given in \cite{Byrnes:2008zy}). In the case that $\fnl$ follows an exact power law behavior, $\fnl\propto k^{\nfnl}$ then $\fnl$ and $\nfnl$ are of course independent. 
However if $\fnl=A\ln(k)+B$ where $A$
and $B$ are constants, then $\nfnl=A/(A\ln(k)+B)=A/\fnl$. In this case $\nfnl$ and $\fnl$ are not independent. Nonetheless one can easily
check that the running of $\nfnl$ satisfies $\alpha_{\fnl}=-\nfnl^2$
so it is a good approximation to treat $\nfnl$ as constant provided
that $|\nfnl|\ll1$.

Consider, as a first example, the curvaton scenario \cite{curvaton}
in the pure curvaton limit. In this case all of the non-Gaussianity
parameters will have some scale dependence unless the curvaton has
exactly a quadratic potential, in which case it can be treated as a
free test field during inflation. This is manifest from eqs.~
(\ref{ntnlcu}) and (\ref{ngnlcu})\footnote{For a quadratic model
$N_{\sigma\sigma\sigma}=0$ (when working to first order in
$r=\rho_{\sigma}/(3H^2)$, i.e. considering the curvaton as a test
field) and hence $g_{\rm NL}=0$. As explained in
Sec.~(\ref{sec:general}), in this case we define $\ngnl=0$ instead
of using the formally divergent result (\ref{ngnlcu}).}. In
\cite{Byrnes:2009pe} we computed $\nfnl$ for curvaton models with a
quartic self-interaction term, $V=m^2\sigma^2/2+\lambda\sigma^4$,
finding a scale dependence proportional to $\eta_{\sigma}$, which
tends to be too small to be of observable interest. The result might
be different for other type of interactions and it would be
interesting to compute $\nfnl$ and $\ngnl$ for generic interacting
curvaton models. This, however, requires a numerical study and is
beyond the scope of the current work \cite{Byrnes:2010xd}. Here we will instead consider
the modulated reheating scenario as an example of isocurvature
single field models. In this case there is little constraint on the
form of the modulaton potential and we can use results derived in
the literature to compute the scale dependencies.

\subsubsection{Modulated reheating}

In this scenario, an isocurvature field $\sigma$ during inflation 
modulates the decay rate of the inflaton field into radiation.
Because the expansion rate of the universe changes after the decay,
this process can convert the initial isocurvature perturbations of
the modulaton field into the primordial curvature perturbation
\cite{modulatedreheating,Suyama:2007bg}. This is closely related to
the model of modulated preheating \cite{modulatedpreheating} and modulated trapping \cite{Langlois:2009jp} (see also \cite{Barnaby:2010ke}). This
process leads to some level of non-Gaussianity, which depends on the
efficiency of the transfer, on the functional form of the decay rate
$\Gamma(\sigma)$ and on the potential of the modulaton field
$V(\sigma)$. The form of the inflaton potential during horizon
crossing is unconstrained, assuming the inflaton perturbations can
be neglected, but its shape around the minimum does influence
reheating and we assume it has a quadratic potential while it is
oscillating.

For simplicity we will consider the case that $\Gamma\ll H_e$, where
$H_e$ is the Hubble parameter measured at the end of inflation
$t_e$. Hence we are assuming that the inflaton decays long after the
end of inflation. In this case, the curvature perturbation in real
space can be written as \cite{Suyama:2007bg,Ichikawa:2008ne}
  \beq
  \label{modulated_zeta}
  \zeta(t_{f},\x)\simeq-\frac{1}{6}\frac{\Gamma_{\sigma_{i}}}{\Gamma}\,\delta\sigma(t_{i},\x)
  +\frac{1}{2}\left(-\frac{1}{6}\frac{\Gamma_{\sigma_{i}}}{\Gamma}\right)_{\sigma_{i}}\times
  \delta\sigma(t_{i},\x)^2+
  \frac{1}{6}\left(-\frac{1}{6}\frac{\Gamma_{\sigma_{i}}}{\Gamma}\right)_{\sigma_{i}\sigma_{i}}\times
  \delta\sigma(t_{i},\x)^3
  +\cdots\ ,
  \eeq
where $t_i$ is a time soon after the horizon crossing of modes of
interest. Using Eq.~(\ref{fnlgnl_pivot}) we find the constant parts
of $\fnl$ and $\gnl$ are given by
  \baq
  \label{mr:fnl}
  \fnl&=&5\left(1-\frac{\Gamma\Gamma_{\sigma_i\sigma_i}}{\Gamma_{\sigma_i}^2}\right)\ ,
  \\
  \label{mr:gnl}
  \gnl&=&
  \frac{50}{3}\left(2
  -3\frac{\Gamma\Gamma_{\sigma_i\sigma_i}}{\Gamma_{\sigma_i}^2}
  +\frac{\Gamma^2\Gamma_{\sigma_i\sigma_i\sigma_i}}{\Gamma_{\sigma_i}^3}\right)\
  .
  \eaq

The scale dependencies of $\fnl$ and $\gnl$ can be computed using
Eqs.~(\ref{ntnlcu}) and (\ref{ngnlcu}). From these equations it is
obvious that a potentially large scale dependence, accompanied with
large values for $\fnl$ and $\gnl$, can arise only if the modulaton
field $\sigma$ has large self interactions. For the rest of this
section we will consider the case of a quartic potential
  \beq
  V(\sigma)=\frac{\lambda}{4!}\sigma^4\ , \qquad \lambda>0\ .
  \eeq
In keeping with the previous literature, we neglect the energy
density of the $\sigma$ field after the end of reheating, as
studying this goes beyond the realms of this project.

Using Eqs.~(\ref{ntnlcu}) and (\ref{ngnlcu}) we find
  \baq
  \label{nfnlmr}
  \nfnl&\;\simeq\;&-\frac{5}{\fnl}\frac{\Gamma}{\Gamma_{\sigma_i}}\frac{\lambda\sigma_i}{3H_i^2}
  \;\sim\; 0.1 \frac{\sqrt{\lambda\eta_{\sigma}}}{{f_{\rm NL}\,{\cal
  P}_{\zeta}^{1/2}}}
  \ ,\\
  \label{ngnlmr}
  \ngnl&\;\simeq\;&2\frac{\fnl^2}{\gnl}\nfnl+\frac{50}{3\gnl}\frac{\Gamma^2}{\Gamma_{\sigma_i}^2}\frac{\lambda}{3H_i^2}
  \;\sim\; 2\frac{\fnl^2}{\gnl}\nfnl+4\times10^{-3}\frac{\lambda}{\gnl\,{\cal
  P}_{\zeta}}\ ,
  \eaq
where $\eta_\sigma = \lambda \sigma_i^2/(6 H_i^2)$. In the
expression for $\ngnl$ we have neglected the contribution
proportional to $\eta_{\sigma}^2$ in Eq.~(\ref{ngnlcu}) which is
negligible compared to $\lambda/(3H_i^2)$ because
$\lambda\sigma_i^{4}\ll H_i^2$ by construction.

The quantities $\nfnl$ and $\ngnl$ could  be large, by making  a
suitable  choice of the parameters. At first sight, it seems easy to
obtain values for these parameters  of order $10^{-1}$, large enough
to be detectable, and at the same time compatible with the
assumptions that underlie our analysis of Section \ref{sec:general}.
This is correct, but we have to ensure that the parameters
  satisfy stringent  constraints
 in order to obtain acceptable values for
the tilt of the spectral index.
 Indeed, assuming inflation lasted
considerably longer than 60 efoldings, a natural initial value for the
 field $\sigma$
 is \cite{Starobinsky:1994bd}
\bea \sigma_i\sim \left(\frac{3}{\pi^2}\right)^{\frac14}\,
 \frac{H_i}{\lambda^{1/4}}\ ,\eea
(a different argument changes the power of $\lambda$ from $1/4$ to
$1/3$ and the numerical factors \cite{Dimopoulos:2003ss}, but the
difference is not very important here). Plugging the previous
estimate in the expression for $\eta_\sigma = \lambda \sigma_i^2/(6
H_i^2)$, and requiring that this parameter is less than $10^{-2}$,
we find the following bound for the coupling $\lambda$: \bea
\eta_{\sigma}\,\simeq\, \left(\frac{\lambda}{12
\pi^2}\right)^{1/2}\lesssim10^{-2}\;\Rightarrow\;
\lambda\lesssim10^{-2}\ . \eea
The condition $\Gamma\ll H_e$ can place further bounds on
$\eta_{\sigma}$ since the modulaton is assumed to remain nearly
frozen until the inflaton decay. We will not further address this issue here.

Plugging the previous results in  (\ref{nfnlmr}) and (\ref{ngnlmr}),
and using ${\cal P}_\zeta = 2.5 \times 10^{-9}$ for the
normalization of the power spectrum,  we find
  \bea
  \left| \nfnl \right|&\;\sim\;&
  \lambda^{3/4}\,\frac{600}{\left| \fnl\right|}\;\lesssim\;\frac{20}{\left| \fnl\right|}\ ,\\
  \label{ngnlestimate}
  |\ngnl|&\;\sim\;&\lambda\,\frac{2\times 10^6}{|\gnl|}\;\lesssim\; \frac{2\times 10^4}{|\gnl|}\
  .
  \eea
where the inequalities are saturated for $\lambda \sim 10^{-2}$. In
the estimate for $\ngnl$, we have neglected the first term in
Eq.~(\ref{ngnlmr}),
  \beq
  \label{mr_ngnl_first}
  2\frac{\fnl^2}{|\gnl|}\,|\nfnl|\;\sim\;
  \lambda^{3/4}\,\frac{10^{3}\,|\fnl|}{|\gnl|}\ ,
  \eeq
which is subdominant compared to the second term if
$\lambda^{1/4}\gtrsim 5\times 10^{-4}|\fnl|$. For $|\fnl|\sim 100$,
this corresponds to $\lambda\gtrsim 6\times 10^{-6}$. In the
opposite case, $\lambda^{1/4}\lesssim 5\times 10^{-4}|\fnl|$, the
estimate for $\ngnl$ is given by Eq.~(\ref{mr_ngnl_first}) instead
of Eq.~(\ref{ngnlestimate}).

We conclude that both $\nfnl$ and $\ngnl$ could acquire relatively
large values, even if the values of $\fnl$ and $\gnl$ saturate their
current observational bounds. It is however important to emphasize
that $|\nfnl|$ or $|\ngnl|\gg 0.01$ is outside the regime of
validity of our formulae, since the accuracy of the expansions
performed in Sec.~\ref{sec:general} starts to become inadequate.


\subsection{Two-field local case}\label{sec:2field}

As a last example, we briefly discuss the so called
 two field local case, for
which $\fsig$ and $\gsig$ are independent of $k$.
This demonstrates an explicit violation of the relation (\ref{crscng}).
 In this scenario, the formulae at the beginning
of this section provide
\bea \tau_{\rm NL}=\left(\frac65\fnl\right)^2\frac{1}{w_{\sigma}}\ ,
\eea
which shows that, in principle, the parameter
$w_{\sigma}$ is an observable. The scale dependences of the
non-linearity parameters satisfy the following relation
\bea \ntaunl=\ngnl=\frac32\nfnl\ . \eea So, as previously stated, we have
a different consistency relation between $\nfnl$ and $\ntaunl$ in
this case compared to the single field case, Eq.~(\ref{crscng}).
Furthermore there is an additional consistency relation from
$\ngnl$. However two-field local models are likely to arise from a
test field with a quadratic potential \cite{Byrnes:2009pe}, in which
case the amplitude of $\gnl$ tends to be too small to be observable.

As an explicit example we consider the mixed inflaton-curvaton
scenario, assuming the curvaton field has a quadratic potential. We
discussed this model previously at the level of the bispectrum in
\cite{Byrnes:2009pe}, and found that in a natural limit
$\nfnl=-2(n_{\zeta}-1)$. It therefore follows that for this model
$\ntaunl$ is even larger,
\bea \ntaunl=-3(n_{\zeta}-1)\ . \eea
In this model $\gnl\sim\fnl$ \cite{Sasaki:2006kq} which is too small to be of observational interest \cite{Smidt:2010ra}.

\section{Shape dependence}\label{sec:shapedependence}

In the previous sections, we concentrated our analysis on the  scale
dependence of equilateral figures (triangles and quadrilatera).
Moreover, we only considered the possibility of varying
simultaneously all of the sides of the figure by the same proportion.
 In this section, we study more general situations in which
scale dependence can arise in parameters characterizing local
non-Gaussianity.  In particular, we consider the case in which the
figure under consideration is not equilateral, and the case in which
we vary the size of only one side, keeping the lengths of the other sides fixed.

We start by
studying these issues for the parameter $f_{\rm NL}$, generalizing the arguments developed in Sec.~\ref{sec_tpf}, and  using the same quantities
 introduced there.
Expanding Eq.~(\ref{fnlsd}),
 around a pivot
scale $k_{\rm p}$ using Eq.~(\ref{fexp}), we obtain
  \baq
  \label{fnl_exp}
  \fnl(k_1,k_2,k_3)&=&\sum_{ab}f_{\rm
  NL}^{ab}\left(\rule{0pt}{4ex}\right.1+\frac{
  k_3^3\left(n_{{\rm multi},a}\ln\,\frac{k_1k_2}{k_{\rm p}^2}+n_{f,ab}\ln\,\frac{k_3}{k_{\rm p}}\right)+2\,{\rm
  perms}}{k_1^3+k_2^3+k_3^3}\left)\rule{0pt}{4ex}\right.\ .
   \eaq
As  in the previous sections, the result is given to first order in
${\rm ln}(k_i/k_{\rm p})$ and both the scale-dependent and
scale-independent parts are given to leading order in slow roll.

In this approximation, setting for simplicity $k_{\rm p}=1$,
equation (\ref{fnl_exp}) can be re-expressed in a more elegant way
as
  \be\label{gensdfnlm} \fnl(k_1, k_2, k_3)\,=\,
  \sum_{ab} f_{\rm NL}^{ab}\frac{\left(k_1 k_2\right)^{n_{{\rm multi}, a}}
  k_3^{3+n_{f,ab}}+2\,{\rm perms}}{k_1^3+k_2^3+k_3^3}\ .
  \ee
Indeed, since both $n_{{\rm multi}, a}$ and $n_{f,ab}$, for each $a,
b$, are proportional to slow-roll parameters,  an expansion of
Eq.~(\ref{gensdfnlm}) at first order in slow-roll provides
Eq.~(\ref{fnl_exp}). For general single field models, it reduces to
\be\label{gensdfnlssa} \fnl(k_1, k_2, k_3)\,=\, \fnl^{\rm p}
 \frac{  k_1^{3+n_{\fnl}}
+ k_2^{3+n_{\fnl}} + k_3^{3+n_{\fnl}}}{k_1^3+k_2^3+k_3^3}\ , \ee
where $\fnl^{\rm p}$ denotes $5f_{\sigma\sigma}/3$ evaluated at the
pivot scale and $n_{\fnl}=n_{f, \sigma\sigma}$. These simple ways of
expressing the parameter $\fnl$ are particularly suitable to analyze
how the triangle shape affects the scale dependence. The single
field expression (\ref{gensdfnlssa}) is equivalent to the analogous
result given in Section 3.3 of \cite{Byrnes:2009pe}, as one can
easily check using Appendix \ref{App:B}. Eq. (\ref{gensdfnlssa})
however takes a much simpler form than the result in
\cite{Byrnes:2009pe} as a consequence of cancellations that occur
when explicitly writing out the results in terms of slow roll
parameters.

We note that, although (\ref{gensdfnlssa}) is not of the form $\fnl\propto (k_1 k_2 k_3)^{\nfnl/3}$ which \cite{Sefusatti:2009xu} used in order to make observational forecasts for $\nfnl$, the bispectrum is a sum of three simple, product separable terms
\bea\label{Bs} B_{\zeta}(k_1,k_2,k_3)\propto
(k_1k_2)^{n_{\zeta}-4}k_3^{\nfnl}+2\;{\rm perms}\ , \eea
and that it only depends on one new parameter $\nfnl$. In the multiple field case the bispectrum will typically depend on more parameters than just $\nfnl$, see (\ref{gensdfnlm}).
An exception is the two-field local model discussed in
 Sec.~\ref{sec:2field}, in which case (we also use Eq.~(\ref{defnfnl2fm}))
\bea\label{Bm} B_{\zeta}(k_1,k_2,k_3)\propto (k_1k_2)^{ n_{\zeta}+(\nfnl/2)-4}
+2\;{\rm perms}\ . \eea

Notice that it therefore follows from (\ref{Bs}) and (\ref{Bm}) that models with the same $\fnl$ and $\nfnl$ can have different bispectral shapes which generalise in different ways the local shape. It is possible that observations may distinguish between these shapes and that we could therefore learn whether $\nfnl$ arises due to single or multi-field effects (or a combination of the two)\footnote{CB thanks Sarah Shandera for pointing this out.}.

In  Sec.~\ref{sec:general}, we limited our considerations to  the
scale dependence of $\fnl$ for equilateral triangles. On the other
hand, by means of Eq.~(\ref{gensdfnlm}) one can observe that,
considering a common rescaling
 for all the three vectors, say  $\k_i \to \alpha \k_i$,
 our previous results remains valid
{\it regardless} of the triangle shape. Namely, \be \frac{\partial
\ln f_{\rm NL}(\alpha k_1, \alpha k_2, \alpha k_3)}{\partial
\ln{\alpha}}\Big|_{\alpha=1}\,=\, \sum_{ab} f_{\rm NL}^{ab}\left(2
n_{{\rm multi}, a} +n_{f,ab}\right)\ . \ee which is exactly our
previous result.

While the scale dependence of $\fnl$, when simultaneously varying the
triangle sides, does not depend on the triangle shape, there are
other situations in which it does.
  We might indeed  be interested on the scale
dependence of $\fnl$, when  varying the size of  only one of the
triangle sides, keeping the other two fixed (and the triangle closed).  In this case, the result
 does depend on the triangle shape. We focus for
simplicity on the single-field case, for which the analysis is
particularly simple, we do not expect our results to change much when considering multiple fields.
  When varying
$\k_1\to \alpha \k_1$, equation (\ref{gensdfnlm})
  becomes
\be\label{gensdfnls} \fnl(\alpha k_1, k_2, k_3)\,=\, \fnl^{\rm p}
 \frac{ \alpha^{3+n_{\fnl}}\, k_1^{3+n_{\fnl}}
+ k_2^{3+n_{\fnl}} + k_3^{3+n_{\fnl}}}{\alpha^3\,k_1^3+k_2^3+k_3^3}\
. \ee  It is clear that the dependence on $\alpha$, in this
expression, goes to zero in the limit in which $k_1$ vanishes. This
is because
 the
coefficients of the terms depending on $\alpha$, in equation
(\ref{gensdfnlm}), become very suppressed with respect to the
remaining  terms.
 This  situation corresponds
to  a squeezed triangle:
 for this shape,
  we then learn that  the value of $\fnl$ does not change
when varying the length of the triangles  shortest side.

In order to determine the triangle shape that leads to  maximal
scale dependence, one is then lead to  focus on the opposite limit.
That is, on
 configurations
for which $k_1^3$ is as large as possible, with respect to
$k_2^3+k_3^3$.  In this case, indeed, the coefficients of the terms
depending on $\alpha$, in equation (\ref{gensdfnlm}), become
dominant with respect to the other terms.

\begin{figure}[!h]
 \begin{center}
 \includegraphics[width=6 cm, height=5  cm]{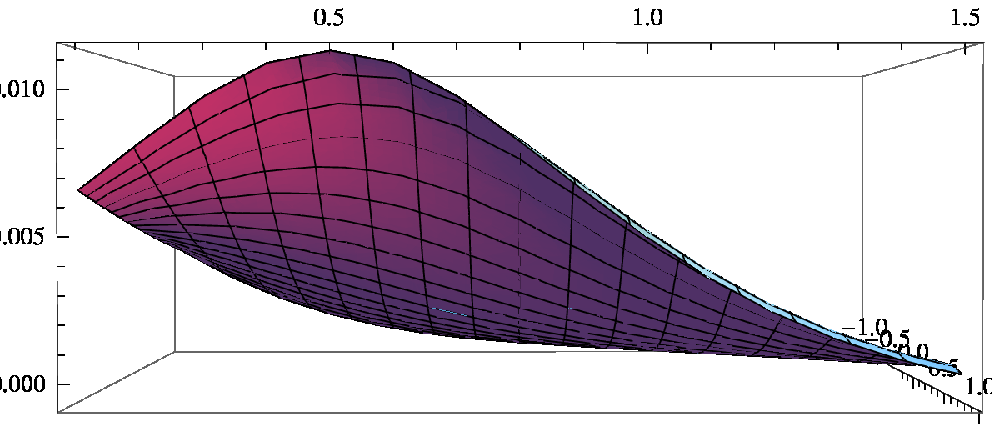}
\hskip 1cm
 \includegraphics[width=6 cm, height=5  cm]{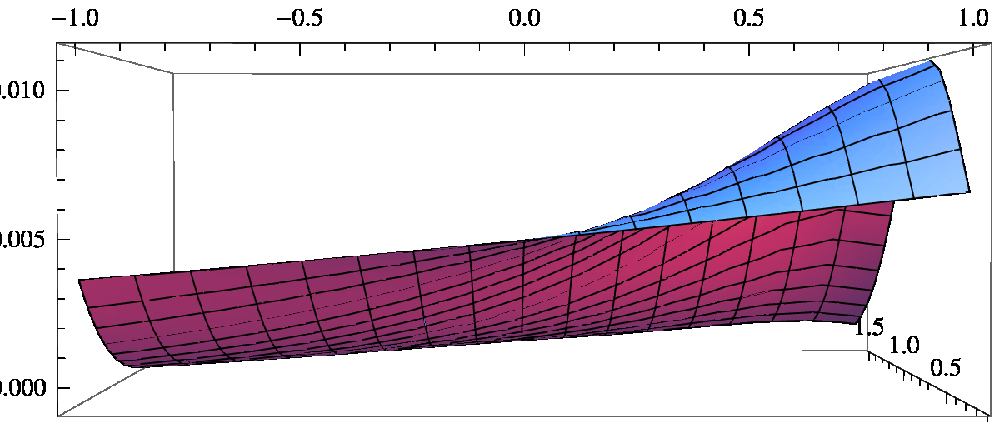}\label{fig_der}
\caption{Behavior of the quantity
$\partial\,\ln{\fnl}/ \partial \ln{\alpha}$,
as a simultaneous function of $x$ (taken between $0$ and $1.5$) and
of $\cos{\theta}$. The two plots represent the same figure from two
different points of view, that emphasize respectively the dependence
on $x$ and on $\cos{\theta}$. We have chosen $\nfnl=0.01$.}
\end{center}
\end{figure}

This expectation is correct, as shown by the following more detailed
analysis. Taking the logarithmic
 derivative of $\fnl$
along $\alpha$, we find, at leading order in slow-roll:
\be\label{derlul} \frac{\partial \ln \fnl(\alpha k_1,  k_2,
 k_3)}{\partial \ln{\alpha}}\Big|_{\alpha=1}
\,=\,\frac{n_{\fnl}}{1+x^3+y^3} \left( 1-\frac{3 x^3 \ln{x}+3y^3
\ln{y}}{1+x^3+y^3} \right)\ , \ee where we have defined $x=k_2/k_1$,
$y=k_3/k_1$. We have checked that the terms inside the parenthesis
are not important for determining the location of the maxima of the
previous expression. The maxima of Eq.~(\ref{derlul}) are therefore
determined by the prefactor $(1+x^3+y^3)^{-1}$, which is maximized
for triangles that minimize the combination $x^3+y^3$. This
corresponds, as anticipated from our previous expectation, to the
shape for which $k_1^3$ is as large as possible, with respect to the
combination $k_2^3+k_3^3$.
  Calling $\theta$ the angle between $k_1$ and $k_2$, we have
$ y^{2}\,=\,\left(1-x\right)^2+2 x \left(1-\cos{\theta}\right) $. So
we can write \be x^3+y^3\,=\,x^3+\left[\left(1-x\right)^2+2 x
\left(1-\cos{\theta}\right) \right]^\frac32\ . \ee It is easy to see
that the previous expression is minimized  for $\theta=0$ and
$x=1/2$, that is for a folded triangle for which $k_2=k_3=k_1/2$.
Plugging these values in (\ref{derlul}), we find
  \be\label{derlul2}
  \frac{\partial \ln \fnl }{\partial \ln{\alpha}}
  {\Big|_{\alpha=1}} \,\simeq\, 1.1\, n_{\fnl}\ ,
  \ee
so we learn that,  for the shape that maximizes the scale
dependence,  we gain around ten per cent with respect to the case in
which we vary simultaneously all the sides of the triangle.
 Plots in Fig. \ref{fig_der}  represent the logarithmic
derivative of $f_{\rm NL}$ along $\alpha$, and graphically show the
results discussed so far. Notice that the shape which maximizes
the scale dependence is indeed
 given by  folded triangles.

A similar procedure, that generalizes what we have done in
Sec.~\ref{sec_qpf}, can be applied to analyze $\gnl$ and $\taunl$.
Expanding Eqs.~(\ref{tnlsd}) and (\ref{gnlsd}) around a pivot scale
$k_{\rm p}$ using Eqs.~(\ref{fexp}), (\ref{gexp}) and (\ref{Pab}),
we obtain the results
  \baq
  \label{gnl_exp}
  \gnl(k_1,k_2,k_3,k_4)&=&\sum_{abc}g_{\rm
  NL}^{abc}\left(\rule{0pt}{4ex}\right. 1+
  \frac{k_4^3\left(n_{{\rm multi},a}\ln\,\frac{k_1k_2k_3}{k_{\rm p}^3}+n_{g,abc}\ln\,\frac{k_4}{k_{\rm
  p}}\right)+4\,{\rm
  perms}}{k_1^3+k_2^3+k_3^3+k_4^3}\left)\rule{0pt}{4ex}\right.\ ,
  \\
  \taunl(k_1,k_2,k_3,k_4,k_{13})&=&\sum_{abcd}\tau_{\rm NL}^{abcd}\left(\rule{0pt}{4ex}\right.\delta_{bc}+
  \left(\rule{0pt}{4ex}\right.
  \frac{\delta_{bc}\left(
  n_{{\rm multi},a}\ln\,\frac{k_1k_2}{k_{\rm p}^2}+
  (n_{bb}-n_{\zeta})\ln\,\frac{k_{13}}{k_{\rm p}}\right)}{k_1^{3}k_2^{3}k_{13}^{3}}
  \qquad\nonumber\\
  &&+
  \frac{\delta_{bc}\left(n_{f,ab}\ln\,\frac{k_3k_4}{k_{\rm
  p}^2}\right)-(1-\delta_{bc})(4\sqrt{\epsilon_b\epsilon_c}
  -2\eta_{bc})\ln\,\frac{k_{13}}{k_{\rm p}}}{k_1^{3}k_2^{3}k_{13}^{3}}+11\,{\rm
  perms}\left)\rule{0pt}{4ex}\right.\qquad\nonumber\\
  &&
  \times
  \left(
  \frac{1}{k_1^{3}k_2^{3}k_{13}^{3}}+11\,{\rm
  perms}\right)^{-1}\left)\rule{0pt}{4ex}\right.\ .\label{taunl_exp}
  \eaq

We can then proceed with arguments very similar to the ones
developed for $\fnl$. Writing $\k_i\rightarrow \alpha\k_i$ in
Eqs.~(\ref{gnl_exp}) and (\ref{taunl_exp}), taking a logarithmic
derivative with respect to $\alpha$ and finally setting $\alpha=1$,
we immediately recover the results (\ref{ntaunl}) and (\ref{ngnl}),
derived in Sec.~\ref{sec:general} for $\ntaunl$ and $\ngnl$ for
equilateral configurations. This shows that for the class of shape
preserving variations, $\k_i\rightarrow \alpha\k_i$, the results are
independent of the figure shape.

In the single field case, if we vary only one of the sides, say the
one labeled by $k_1$, then the scale dependence vanishes when $k_1
\to 0$.   We numerically analyzed for which shapes the scale
dependence is maximal. For the case of $g_{\rm NL}$, the analysis is
a straightforward generalization of what we did for $\fnl$. The
shape associated with maximal scale dependence corresponds to a
folded polygon, in which three of the sides lie over the side whose
length is varied. That is,
 \be \label{folq} k_1= 3 k_2 = 3 k_3 =3 k_4\ .\ee Again, for maximal
scale dependence we gain order ten per cent with respect to the case
in which we vary all the sides simultaneously.

 We also performed a numerical
analysis to study $\taunl$, finding again maximal scale dependence
for the folded shape of Eq.~(\ref{folq}). For this parameter, we
gain around $20-25$ percent
   with respect
to the case in which we vary all the sides by the same amount (this
result resonates with the consistency relation (\ref{crscng})).

\section{Curvature perturbation in coordinate space}\label{sec:coordspace}

An additional important feature of our approach to the
  scale dependence of local non-Gaussianity, is that it allows one
 to express the results in coordinate space.
 In this section, we show how the scale dependent coefficients
appearing in the momentum space expansion of curvature perturbation
(\ref{zeta_k_3rd}) manifest themselves in coordinate space. It is
clear that scale dependence will cause deviations from the local
form, for which $\zeta(\x)$ can be expressed as a power series of a
Gaussian variable $\zeta^{\G}(\x)$ with constant coefficients. Using
the general single field case as an example, we work out the
expression for $\zeta(\x)$ in the scale-dependent case and quantify
how it deviates from the local form.

In the general single field case, Eq.~(\ref{zeta_fg_1}) can be
written as
  \baq
  \label{zeta_k_coord}
  \zeta_{\k}&=&\zeta_{\k}^{\G}+\frac{3}{5}\fnl(k_{\rm p})\left(1 +\nfnl\,\theta(k_i-k)\,{\rm ln}
  \frac{k}{k_i}\right)
  (\zeta^{\G}\star\zeta^{\G})_{\k}\\\nonumber&&+
  \frac{9}{25}\gnl(k_{\rm p})\left(1 +\ngnl\,\theta(k_i-k)\,{\rm ln}
  \frac{k}{k_i}\right)(\zeta^{\G}\star\zeta^{\G}\star\zeta^{\G})_{\k}+\cdots\
  ,
  \eaq
which coincides with Eq.~(\ref{zeta_single}) up to slow roll
corrections for constant terms. This form is useful for our analysis
since the horizon scale $k_i=a_iH_i$ appears explicitly. We have
inserted the theta functions $\theta(k_i-k)$ to explicitly indicate
that the result holds only for super-horizon modes $k< k_i$. $k_i>k_p$ should correspond to a physically smaller scale than any of the modes of interest.  Recall
that similar theta functions are included in our definition of
$\zeta_{\k}^{\G}$, Eq.~(\ref{zeta_G}). Therefore $\zeta_{\k}^{\G}$
can be viewed as a smoothed quantity; in Fourier space the window
function is simply a top hat with the cutoff set at the horizon
scale $k_i$.

Taking the inverse Fourier transform of (\ref{zeta_k_coord}) we find
  \baq
  \label{zeta_coord_1}
  \zeta(\x)&=&\zeta^{\G}(\x)+\frac{3}{5}\fnl(k_{\rm p})\zeta^{\G}(\x)^2+
  \frac{9}{25}\gnl(k_{\rm p})\zeta^{\G}(\x)^3\\\nonumber
  &&+\frac{3}{5}\fnl(k_{\rm p})\nfnl\int\frac{{\rm d}{\bf
  k}}{(2\pi)^3}e^{i
  \k\cdot\x}\theta(k_i-k)(\zeta^{\G}\star\zeta^{\G})_{\k}\,{\rm
  ln}\frac{k}{k_i}\\\nonumber
  &&+\frac{9}{25}\gnl(k_{\rm p})\ngnl\int\frac{{\rm d}{\bf
  k}}{(2\pi)^3}e^{i
  \k\cdot\x}\theta(k_i-k)(\zeta^{\G}\star\zeta^{\G}\star\zeta^{\G})_{\k}\,{\rm
  ln}\frac{k}{k_i}+\cdots\ .
  \eaq
The two integrals describe deviations from the local form. They can
be written more explicitly by performing the following manipulations
  \baq
  \label{def_I}
  \int\frac{{\rm d}{\bf
  k}}{(2\pi)^3}e^{i
  \k\cdot\x}\theta(k_i-k)(\zeta^{\G}\star\zeta^{\G})_{\k}\,{\rm
  ln}\frac{k}{k_i}&=&\int{\rm d}{\bf y}\int\frac{{\rm d}{\bf
  k}}{(2\pi)^3}e^{i
  \k\cdot(\x-{\bf y})}\theta(k_i-k)\,\zeta^{\G}({\bf y})^2{\rm
  ln}\frac{k}{k_i}\qquad\\\nonumber&=&
  \int{\rm d}{\bf y}\zeta^{\G}({\bf y})^2 \frac{1}{2\pi^2}
  \frac{{\rm sin}(k_i |\x-{\bf y}|)-{\rm Si}(k_i |\x-{\bf y}|)}{|\x-{\bf y}|^3}
  \\\nonumber&\equiv&
  \int{\rm d}{\bf y}\zeta^{\G}({\bf y})^2I(|\x-{\bf y}|)\ ,
  \eaq
and similarly for the second integral. Using this we can rewrite
Eq.~(\ref{zeta_coord_1}) as
  \baq
  \label{zeta_coord_2}
  \zeta(\x)&=&\zeta^{\G}(\x)+
  \frac{3}{5}\fnl(k_{\rm p})\left(\zeta^{\G}(\x)^2+\nfnl\int{\rm d}{\bf y}I(|\x-{\bf y}|)\zeta^{\G}({\bf y})^2\right)
  \\\nonumber
  &&+
  \frac{9}{25}\gnl(k_{\rm p})\left(\zeta^{\G}(\x)^3
  +\ngnl\int{\rm d}{\bf y}I(|\x-{\bf y}|)\zeta^{\G}({\bf
  y})^3\right)+\cdots\ .
  \eaq

\begin{figure}[h!]
    \begin{center}
    \includegraphics[width=12 cm, height= 3.6 cm]{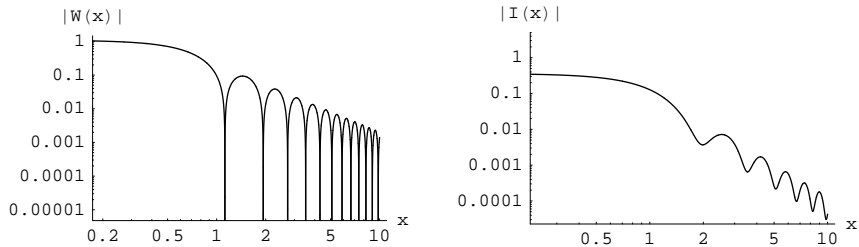}
    \label{W_and_I_W}
    \caption{Absolute values of the functions $W$ and $I$ plotted on logarithmic scales for the choice $k_i=4$
    (in arbitrary units).}
    \end{center}
  \end{figure}

This result clearly shows how the scale dependence of
$f_{\sigma\sigma}$ and $g_{\sigma\sigma\sigma}$ in
Eq.~(\ref{zeta_k_3rd}) renders $\zeta(\x)$ a nonlocal function of
$\zeta^{\G}(\x)$. Because of the integrals in
Eq.~(\ref{zeta_coord_2}), the curvature perturbation $\zeta(\x)$ can
not be expressed in terms of $\zeta^{\G}(\x)$ evaluated at the same
point $\x$ but one needs to know $\zeta^{\G}({\rm y})$ in the entire
region where $I(|\x-{\bf y}|)$ is non-vanishing. The behavior of
$I(|\x-{\bf y}|)$ is depicted in Fig.~\ref{W_and_I_W} which also
displays the inverse Fourier transform, 
\bea W(|\x-{\bf y}|)=\frac{{\rm
sin}(k_i\,|\x-{\bf y}|)-k_i\,|\x-{\bf y}|\,{\rm cos}(k_i\,|\x-{\bf
y}|)}{2\pi^2|\x-{\bf y}|^3},   \eea
 of the top hat window function
$\theta(k_i-k)$ included in the definition of $\zeta^{\G}_{\k}$,
Eq.~(\ref{zeta_G}).
Both $W(x)$ and $I(x)$ are approximatively constant at scales $x
\lesssim (k_i)^{-1}$. They both fall off for $x\gtrsim (k_i)^{-1}$
but $I(x)$ remains negative definite unlike $W(x)$ which starts to
oscillate rapidly. Keeping in mind that a smoothing over $W(x)$ is
implicit in the definition of $\zeta^{\G}(\x)$, we therefore see
that the convolutions of $\zeta^{\G}(\x)$ with $I(x)$ in
Eq.~(\ref{zeta_coord_2}) pick up a non-trivial contribution from the
superhorizon modes $x\gtrsim (k_i)^{-1}$ where $W(x)$ effectively
falls off faster than $I(x)$. This contribution makes
Eq.~(\ref{zeta_coord_2}) deviate from the local form.

The analysis can in principle be straightforwardly generalized to
the multi-field case. The difference compared to the general single
field case is the appearance of several unrelated Gaussian fields
$\zeta^{\G,a}_{\k}$ in Eq.~(\ref{zeta_k_3rd}). This in general makes
it impossible to write $\zeta(\x)$ as a series of a single Gaussian
field even if the coefficients in Eq.~(\ref{zeta_k_3rd}) would be
constants.

\section{Conclusions}\label{sec:conclusions}

We have discussed  a new approach, based on
the  $\delta N$-formalism, for studying the scale dependence of
   non-Gaussianity parameters.
   We have obtained explicit expressions
for the scale dependence of the quantities $\fnl$,
$\taunl$ and $\gnl$ associated with the bispectrum
  and trispectrum of primordial  curvature perturbations.
   Our results depend
on the slow-roll parameters evaluated at horizon exit, and on the
derivatives of the number of  e-foldings and  the inflationary potential.
  The parameters  controlling the scale dependence
  of non-Gaussianity depend on
  properties of the
   the inflationary potential, namely its third and fourth
derivatives,
   which in all observationally interesting cases cannot
be probed by only studying the spectral index of the power spectrum
and its running.

As a consequence,  the scale dependence of non-Gaussianity provides
additional  powerful observables, able to offer novel information
about the mechanism which generates the curvature perturbations. We
demonstrated these features  in the concrete example of modulated
reheating. In models with a quartic potential for the modulating
field, we have shown that the associated non-linearity
parameters, and their scale dependence, can be large enough to be
observable.

While in most of the discussion we  worked  in momentum space, in
the last part we also discussed how to describe our results in
coordinate space. We provided an expression for curvature
perturbations in coordinate  space, that generalizes the frequently
used local Ansatz, and that exhibits directly in real space the
effects of  scale dependence of non-Gaussian parameters.

Our results allow us to put onto a firm basis the phenomenological
parameterizations of the scale dependence of non-Gaussian
observables. In
many models of observational interest, our formulae are relatively
simple and depend on a single new parameter, the scale dependence of
the non-linearity parameter. It would be interesting to use these
results for analysing or simulating non-Gaussian data. At the same
time, our investigation allows us to identify which properties
inflationary models  have to satisfy, in order to obtain large
non-Gaussianity with sizeable scale dependence. It would be
interesting to apply it to analyse further models, for example
those in which multiple fields interact during inflation or where the non-Gaussianity is generated by an inhomogeneous end of inflation.

\acknowledgments

The authors thank Francis Bernardeau, Vincent Desjacques, Tommaso Giannantonio, Cristiano Porciani, Dragan Huterer, Emiliano Sefusatti and Sarah Shandera for interesting discussions. 
CB and DW thank the organisers of the
workshop ``The non-Gaussian universe'' (YITP-T-09-05), in the Yukawa Institute for Theoretical Physics, Kyoto, Japan for hospitality where this work
was discussed. M.G. acknowledges support from the {\it Studienstiftung des Deutschen Volkes}. SN is partially funded by the Academy of Finland
grant 136600. DW is supported by STFC.

\appendix

\section*{Appendices}

\section{Explicit expressions for $n_{f,ab}$ and $n_{g,ab}$}
\label{App:A}

For a system of slowly rolling scalar fields $\phi_{a}$, the
equations of motion are given by $3 H \dot{\phi}_a = -V_a$ and
$3H^2=V$, to leading order in slow roll. Here we are interested in
the evolution during a short time interval from $t_0$ to $t_i$. The
slow roll equations can easily be solved for $\phi_{a}$ as
  \beq
  \label{app_phi_a}
  \phi^a(t_i) = \phi^a(t_0)-\sqrt{2\epsilon_a}\,{\rm ln}(a_i/a_0)+{\cal O}\left(\epsilon^{3/2}{\rm ln}^2(a_i/a_0)\right)\ ,
  \eeq
where we have used the identity $H\d t=\d {\rm ln}\,a$, which holds
to leading order in slow roll. In the following we use the notation
${\cal O}(\epsilon^{n})$ to denote the combinations of the slow roll
parameters $\epsilon_{a}, \eta_{ab}$ of order $\epsilon^{n}$.

Differentiating Eq.~(\ref{app_phi_a}) with respect to
$\phi^{a}(t_0)$ and keeping the number of e-foldings ${\rm
ln}(a_i/a_0)$ fixed, we can compute the coefficients appearing in
Eq.~(\ref{phi_ti}). We choose the initial time $t_{0}$ as the time
$t_{k}$ of horizon crossing of a mode $k$, defined by $a_k H_k=k$.
Using $\ln (a_{i}/a_{k})=\ln
(a_{i}H_{i}/k)$, which is valid at leading order in slow roll, the three first coefficients in Eq.~(\ref{phi_ti})
can be written as
  \baq
  \label{app_coeff_1}
  \frac{\partial\phi^{a}(t_i)}{\partial\phi^b(t_k)}
  &=&
  \delta_{ab}+\epsilon_{ab}\ln\,\frac{a_{i}H_{i}}{k}+{\cal
  O}\left(\left\{\epsilon^{2}, \frac{\epsilon^{1/2}V'''}{3H^2}\right\}\left({\rm ln}\frac{a_iH_i}{k}\right)^2\right)\ ,
  \\
  \label{app_coeff_2}
  \frac{\partial^2\phi^{a}(t_i)}{\partial\phi^b(t_k)\partial\phi^c(t_k)}
  &=&F^{(2)}_{abc}\ln\,\frac{a_{i}H_{i}}{k}+
  {\cal O}\left(\left\{\epsilon^{5/2},\frac{\epsilon V'''}{3H^2},
  \frac{\epsilon^{1/2}V''''}{3H^2}
  \right\}\left({\rm ln}\frac{a_iH_i}{k}\right)^2\right)
  \ ,
  \\
  \label{app_coeff_3}
  \frac{\partial^3\phi^{a}(t_i)}{\partial\phi^b(t_k)\partial\phi^c(t_k)\partial\phi^d(t_k)}
  &=&
  F^{(3)}_{abcd} \ln\,\frac{a_{i}H_{i}}{k}+{\cal
  O}\left(\left\{\epsilon^{3},\frac{\epsilon^{3/2} V'''}{3H^2}
  , \frac{\epsilon V''''}{3H^2},\frac{\epsilon^{1/2} V'''''}{3H^2}\right\}
  \left({\rm ln}\frac{a_iH_i}{k}\right)^2\right)
  \ ,\qquad
  \eaq
where the primes in ${\cal O}(V''')$ etc. denote derivatives with
respect any of the fields $\phi_{a}$, and
  \baq
  F^{(2)}_{abc}
  &=& {\sqrt{2}}\left(-4\sqrt{\epsilon_a\epsilon_b\epsilon_c}+\eta_{ab}\sqrt{\epsilon_c}
  +\eta_{bc}\sqrt{\epsilon_a}+\eta_{ca}\sqrt{\epsilon_b}-\frac{1}{\sqrt{2}}\frac{V_{abc}}{3H^2}\right)\ ,\\
  F^{(3)}_{abcd}
  &=&
 -4(\eta_{ad}\sqrt{\epsilon_b\epsilon_c}+\eta_{bd}\sqrt{\epsilon_c\epsilon_a}+
  \eta_{cd}\sqrt{\epsilon_a\epsilon_b})\\\nonumber&&+
  \sqrt{\epsilon_d}\left(24\sqrt{\epsilon_a\epsilon_b\epsilon_c}-4\eta_{ab}\sqrt{\epsilon_c}
  -4\eta_{bc}\sqrt{\epsilon_a}-4\eta_{ca}
  \sqrt{\epsilon_b}+\sqrt{2}\frac{V_{abc}}{3H^2}\right)\\\nonumber&&+
  \sqrt{2\epsilon_c}\frac{V_{abd}}{3H^2}+\sqrt{2\epsilon_a}\frac{V_{bcd}}{3H^2}+\sqrt{2\epsilon_b}\frac{V_{cad}}{3H^2}+
  \eta_{ab}\eta_{cd}+\eta_{bc}\eta_{ad}+\eta_{ca}\eta_{bd}-\frac{V_{abcd}}{3H^2}\
  .
  \eaq

Substituting these into Eq.~(\ref{dnfab}),
  \baq
  \label{app_nf}
  n_{f,ab}&=&-\sum_{c}\frac{N_c
  F^{(2)}_{cab}}{N_{ab}}\ ,\\
  \label{app_ng}
  n_{g,abc}&=&-\sum_{d}
   \left(3\frac{N_{da}}{N_{abc}}F^{(2)}_{dbc}+\frac{N_{d}}{N_{abc}}F^{(3)}_{dabc}\right)\
   ,
  \eaq
we obtain fully explicit results for the parameters $n_{f,ab}$ and
$n_{g,abc}$. The results are derived retaining only terms up to
first order in $\ln (a_iH_i/k)$ in Eqs.~(\ref{app_coeff_1}) -
(\ref{app_coeff_3}). Higher order terms are suppressed by slow-roll
parameters and their combinations with the derivatives of the
potential. The former are small by construction and the latter also
naturally remain small, provided that the flatness of the scalar
field potential during inflation is not a result of extreme
fine-tuning. Furthermore, since the logarithms never grow very large
for the observable super-horizon modes, $\ln(a_iH_i/k)\lesssim {\cal
O}(10)$, we conclude that the higher order contributions can indeed
be neglected at first order.

\section{On the different formulations of the $\delta N$ approach}
\label{App:B}

The $\delta N$ expression for the super-horizon-scale curvature
perturbation
  \beq
  \label{app_zeta_ti_fourier}
  \zeta_{\k} (t_f)=\sum_{a}N_{a}(t_f,t_i)\delta\phi^{a}_{\k} (t_i)
  +\frac{1}{2}\sum_{ab}N_{ab}(t_f,t_i)
  \int\frac{\d {\bf q}}{(2\pi)^3}\delta\phi^{a}_{\q}(t_i) \delta\phi^{b}_{\k-\q}(t_i) +\cdots\
  ,
  \eeq
is by construction independent of the choice of the initial
spatially flat hypersurface $t_i\geq t_{k}$. This property follows
from the definition of the curvature perturbation \cite{lms} as
demonstrated in \cite{Byrnes:2009pe}. A commonly used choice is to
set $t_{i}$ equal to the time $t_{k}$ of horizon crossing of the
mode $k$. The analysis in \cite{Byrnes:2009pe}
was performed using this choice. In our current work, we have
instead chosen $t_i$ as a time soon after $t_{k}$ following e.g.
\cite{Sasaki:1995aw}. Here we explicitly compare the two choices.
For simplicity, we consider only terms up to second order in
Eq.~(\ref{app_zeta_ti_fourier}). Generalization to higher orders is
straightforward.

Using the chain rule it is easy to switch between the coefficients
$N_{ab..}(t_f,t_i)$ and $N_{ab..}(t_f,t_k)$ in the two different
formulations. For the first and second order terms shown in
Eq.~(\ref{app_zeta_ti_fourier}) we obtain
  \baq
  \label{app_N_1}
  N_{a}(t_f,t_i)&\equiv&\frac{\partial
  N(t_f,t_i)}{\partial\phi^{a}(t_i)}=\sum_{b}\frac{\partial
  N(t_f,t_k)}{\partial\phi^{b}(t_k)}\frac{\partial
  \phi^{b}(t_k)}{\partial\phi^{a}(t_i)}\ ,\\
  \label{app_N_2}
  N_{ab}(t_f,t_i)&\equiv&\frac{\partial^2
  N(t_f,t_i)}{\partial\phi^{a}(t_i)\partial\phi^{b}(t_i)}=\sum_{c}\frac{\partial
  N(t_f,t_k)}{\partial\phi^{c}(t_k)}\frac{\partial^2
  \phi^{c}(t_k)}{\partial\phi^{a}(t_i)\partial\phi^{b}(t_i)}\\\nonumber
  &&
  +\sum_{cd}\frac{\partial
  \phi^{c}(t_k)}{\partial\phi^{a}(t_i)}\frac{\partial
  \phi^{d}(t_k)}{\partial\phi^{b}(t_i)}\frac{\partial^2
  N(t_f,t_k)}{\partial\phi^{c}(t_k)\partial\phi^{d}(t_k)}\ .\qquad
  \eaq
In the second equality of both equations we have replaced the time argument $t_i$ in
$N(t_f,t_i)$ by $t_k$ making use of the fact that $t_i$ and $t_{k}$
both label spatially flat hypersurfaces. This implies that
$N(t_i,t_k)$, the number of e-foldings from $t_k$ to $t_i$, is a
constant under differentiation with respect to the fields.
Writing $N(t_f,t_i)=N(t_f,t_k)-N(t_i,t_k)$, we thus immediately see
that $N(t_f,t_i)$ can be replaced by $N(t_f,t_k)$ in
Eqs.~(\ref{app_N_1}) and (\ref{app_N_2}).

Substituting Eqs.~(\ref{app_N_1}) and (\ref{app_N_2}) into
Eq.~(\ref{app_zeta_ti_fourier}), we obtain
  \baq
  \label{app_zeta_tk_fourier_step1}
  \zeta_{\k} (t_f)&=&\sum_{a}N_{a}(t_f,t_k)\left(\frac{\partial
  \phi^{a}(t_k)}{\partial\phi^{b}(t_i)}\delta\phi^{b}_{\k}(t_i)+\frac{1}{2}
  \frac{\partial^2
  \phi^{a}(t_k)}{\partial\phi^{b}(t_i)\partial\phi^{c}(t_i)}
  \int\frac{\d {\bf q}}{(2\pi)^3}\delta\phi^{b}_{\q}(t_i) \delta\phi^{c}_{\k-\q}(t_i)\right)
  \\\nonumber
  &&+\frac{1}{2}\sum_{ab}N_{ab}(t_f,t_k)\left(\frac{\partial
  \phi^{a}(t_k)}{\partial\phi^{c}(t_i)}\frac{\partial
  \phi^{b}(t_k)}{\partial\phi^{d}(t_i)}
  \int\frac{\d {\bf q}}{(2\pi)^3}\delta\phi^{c}_{\q}(t_i) \delta\phi^{d}_{\k-\q}(t_i)\right)
+\cdots\ .
  \eaq
On the other hand, according to Eq.~(\ref{phi_ti}) we have
  \beq
  \label{app_deltaphi}
  \delta\phi_{\k}^{a}(t_k)=\frac{\partial
  \phi^{a}(t_k)}{\partial\phi^{b}(t_i)}\delta\phi^{b}_{\k}(t_i)+\frac{1}{2}
  \frac{\partial^2
  \phi^{a}(t_k)}{\partial\phi^{b}(t_i)\partial\phi^{c}(t_i)}
  \int\frac{\d {\bf q}}{(2\pi)^3}\delta\phi^{b}_{\q}(t_i)
  \delta\phi^{c}_{\k-\q}(t_i)+\cdots \ .
  \eeq
In arriving at this result we have first taken the Fourier transform
of Eq.~(\ref{phi_ti}) and only thereafter set one of the time
arguments equal to $t_k$. Using Eq.~(\ref{app_deltaphi}) we can
rewrite Eq.~(\ref{app_zeta_tk_fourier_step1}) as
  \baq
  \label{app_zeta_tk_fourier_step2}
  \zeta_{\k} (t_f)&=&\sum_{a}N_{a}(t_f,t_k)\delta\phi_{\k}^{a}(t_k)
  +\frac{1}{2}\sum_{ab}N_{ab}(t_f,t_k)
  \int\frac{\d {\bf q}}{(2\pi)^3}\delta\phi^{a}_{\q}(t_k) \delta\phi^{b}_{\k-\q}(t_k)
  +\cdots\ .
  \eaq
This way of writing $\zeta_{\k}(t_f)$ is equivalent to
Eq.~(\ref{app_zeta_ti_fourier}) and the two expressions differ
formally only by the choice of the initial time $t_i$, as expected.
The relation between the coefficients in the two formulations is
given by Eqs.~(\ref{app_N_1}) and (\ref{app_N_2}), and the field
perturbations are related by Eq.~(\ref{app_deltaphi}). These results
explicitly show how to switch from one formulation to another.

In \cite{Byrnes:2009pe}, the
result for $\nfnl$, measuring the scale dependence of $\fnl$, was
expressed in terms of the parameters $n_{I}={\rm d}\ln
N_{I}(t_f,t_k)/{\rm d}\ln k$ and $n_{IJ}={\rm d}\ln
N_{IJ}(t_f,t_k)/{\rm d}\ln k$, see e.g. Eq.~(69) in that paper.
(Here we follow the notation of \cite{Byrnes:2009pe} and label the
scalar field species $\phi_{I}$ by capital letters. This also serves
to distinguish the parameters $n_I$ and $n_{IJ}$ from the quantities
defined in the current work.) Using Eqs.~(\ref{app_N_1}) and
(\ref{app_N_2}) together with the results derived in Appendix
\ref{App:A}, we readily obtain explicit expressions for these
parameters
  \baq
  n_{I}&=&\frac{{\rm d}\,\ln\, N_{I}(t_f,t_k)}{{\rm d}\,\ln\,k}=-\sum_{J}\frac{N_{J}}{N_{I}}\,\epsilon_{IJ}\ ,\\
  n_{IJ}&=&\frac{{\rm d}\,\ln\, N_{IJ}(t_f,t_k)}{{\rm
  d}\,\ln\,k}=n_{f,IJ}-\sum_{K}\left(\frac{N_{IK}\,\epsilon_{KJ}}{N_{IJ}}
  +\frac{N_{JK}\,\epsilon_{KI}}{N_{IJ}}\right)\ .
  \eaq
In the rightmost expressions we have suppressed the time arguments
$t_i$ for brevity, e.g. $N_{I}\equiv N_{I}(t_f,t_i)$. Using these
results, it is straightforward to check that the general expression
given for $\nfnl$ in Eq.~(69) of \cite{Byrnes:2009pe} agrees with
our Eq.~(\ref{n_fnl_equil}).

\end{document}